%% file: filaments_saga.tex
\documentclass[traditabstract]{aa}  
\usepackage{graphicx}
\usepackage{txfonts}
\usepackage{natbib}
\usepackage{multirow}

\newcommand\micron[0]{\,$\mu$m\xspace}
\newcommand\msun[0]{M$_{\odot}$}

\newcommand{\dzco}{$^{13}$CO\xspace}
\newcommand\hii{H{\sc ii}\xspace}

\begin{document}

   \title{Giant molecular filaments in the Milky Way}


   \author{S.E. Ragan \inst{\ref{mpia}}, Th. Henning \inst{\ref{mpia}}, J. Tackenberg \inst{\ref{mpia}}, H. Beuther \inst{\ref{mpia}}, K.G. Johnston \inst{\ref{mpia}}, J. Kainulainen \inst{\ref{mpia}}, H. Linz \inst{\ref{mpia}} 
          }

   \institute{Max Planck Institute for Astronomy,
              K\"onigstuhl 17, 69117 Heidelberg, Germany\\
              \email{ragan@mpia.de} \label{mpia}
             }

   \date{Received 10 January 2014; accepted 20 June 2014}
   \authorrunning{Ragan et al.}
   \titlerunning{GMFs in the Milky Way}

\abstract{Throughout the Milky Way, molecular clouds typically appear filamentary, and mounting evidence indicates that this morphology plays an important role in star formation.  What is not known is to what extent the dense filaments most closely associated with star formation are connected to the surrounding diffuse clouds up to arbitrarily large scales.  How are these cradles of star formation linked to the Milky Way's spiral structure?  Using archival Galactic plane survey data, we have used multiple datasets in search of large-scale,  velocity-coherent filaments in the Galactic plane.  In this paper, we present our methods employed to identify coherent filamentary structures first in extinction and confirmed using Galactic Ring Survey data.  We present a sample of seven Giant Molecular Filaments (GMFs) that have lengths of order $\sim$100\,pc, total masses of 10$^4$ - 10$^5$\,M$_{\odot}$, and exhibit velocity coherence over their full length. The GMFs we study appear to be inter-arm clouds and may be 
the Milky Way analogues to spurs observed in nearby spiral galaxies.  We find that between 2 and 12\% of the total mass (above $\sim$10$^{20}$ cm$^{-2}$) is ``dense'' (above 10$^{22}$ cm$^{-2}$), where filaments near spiral arms in the Galactic midplane tend to have higher dense gas mass fractions than those further from the arms. 
}


   \maketitle
%

\section{Introduction}

Star-forming clouds in the Milky Way -- both nearby and distant -- exhibit elongated structures \citep[see][and references therein]{Myers2009}. The morphology appears most enhanced when viewing the densest part of the cloud, which is also the regime most intimately connected to star formation. From the early quiescent phase, filamentary morphology seems to be imprinted on all subsequent stages of star formation.  However, the physical origin of filaments is still debated.  A variety of models can produce dense, filamentary structures, though unambiguous observational diagnostics are still lacking to determine the dominant mechanism(s) leading to filament formation. 

Filamentary structures have been observed in a variety of tracers, ranging from extinction maps at optical and infrared wavelengths \citep[e.g.][]{SchneiderElmegreen1979, Apai2005, Jackson2010, Schmalzl2010, Beuther2011, Kainulainen2013} to CO maps \citep{Ungerechts1987, Goldsmith2008, Hacar2013} to far-infrared/sub-millimeter dust emission maps \citep[e.g.][]{Henning2010, Andre2010, Menshchikov2010, Molinari2010, Schneider2010, Hill2011, Hennemann2012, Peretto2012}.  The recent results of {\em Herschel} have again highlighted the ubiquity of filaments in the interstellar medium (ISM) and thus rejuvenated interest in the role filamentary morphology plays in star formation. 

In numerical models, filamentary structure is a natural consequence of a number of dynamic processes in the ISM such as converging flows \citep[e.g.][]{Elmegreen1993, Vazquez-Semadeni2006, Heitsch2008, Clark2012}, the collision of shocked sheets \citep{Padoan2001}, instabilities in self-gravitating sheets \citep[e.g.][]{Nagai1998}, or other analogous processes that compresses gas to an over-dense interface.  Such processes commonly occur in a global spiral potential \citep{Dobbs2008}. In non-self-gravitating cases, additional ingredients, such as magnetic fields and/or turbulence dissipation \citep{Hennebelle2013a}, are needed to preserve filaments with the properties observed in the ISM and in turbulent simulations. More massive filaments, such as those modeled by \citet{Fischera2012} and \citet{Heitsch2013a, Heitsch2013b}, are self-gravitating, thus the effects of continuing accretion from large scales and external pressurisation, play an important role in the observed properties.

A fundamental quantity in the understanding of the origin of Galactic filaments is the maximum length over which they can occur. This is challenging to observe for several reasons. First, until recently, few unbiased surveys of the Galactic plane that could potentially identify such structures existed. Second, the star formation occurring within filaments, especially massive ones, is very disruptive and impacts the clouds and clear signatures of the parent structure. As such, the quiescent stage of filaments, appearing as infrared-dark clouds (IRDCs), should better preserve the initial formation signatures compared to active clouds.

\citet{Jackson2010} identified a long infrared-dark 80\,pc long ``Nessie'' filament. \citet{Goodman2013}\footnote{\url{http://authorea.com/249}} revisited ``Nessie'' finding that it coincides with the Scutum-Centaurus arm, and it may even be at least twice as long. Further searches have found similarly enticing individual structures in the Galactic plane \citep[e.g.][]{Beuther2011,Battersby2012,Tackenberg2013,Li2013}, but to date, no comprehensive compilation of long, coherent structures in the Galaxy exists. In this paper, we present a new sample of Giant Molecular Filaments (GMFs) in the first quadrant of the Milky Way.  We describe our methods for identifying filaments in projection using unbiased Galactic plane surveys and our follow-up method for confirming a filament's coherence in velocity space. Our search has produced seven new filaments with lengths on the order of 100\,pc, more than doubling the number of similar structures known from the literature. This catalog will aid in studying the 
connection between large scale filamentary structure and star formation in the Galaxy.

\begin{figure}
\includegraphics[width=0.5\textwidth]{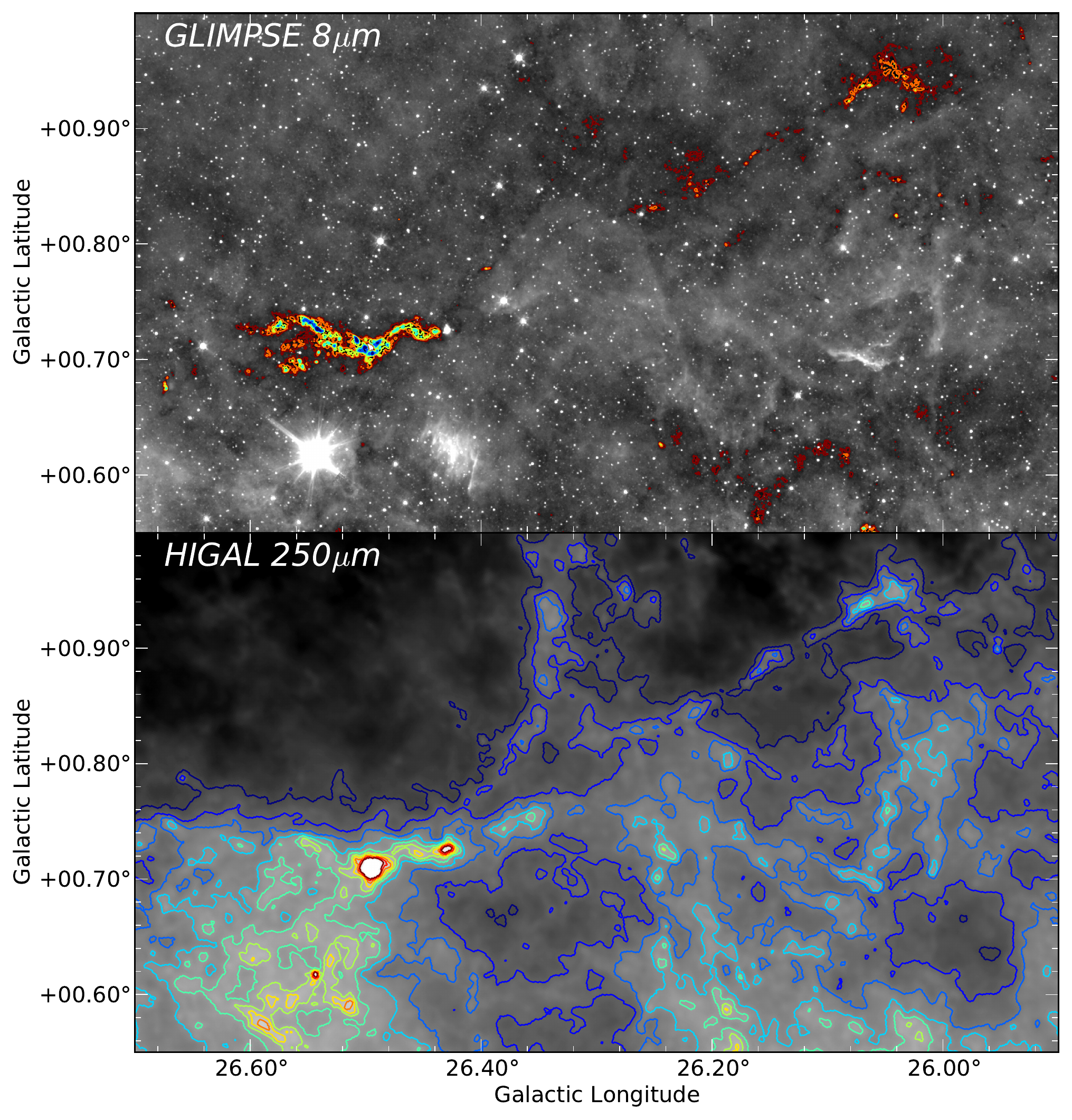}
\caption{\label{fig:higal} Zoom in to the eastern end of the F26.7-25.4 filament.  A Grayscale GLIMPSE 8\micron image (top) is plotted with contours at -7.5, -10, -12.5... MJy sr$^{-1}$ (negative to highlight absorption feature), and the HIGAL 250\micron image (bottom) is plotted with contours drawn at 5, 6, 7... Jy beam$^{-1}$.}
\end{figure}

\begin{figure}
\includegraphics[width=0.53\textwidth]{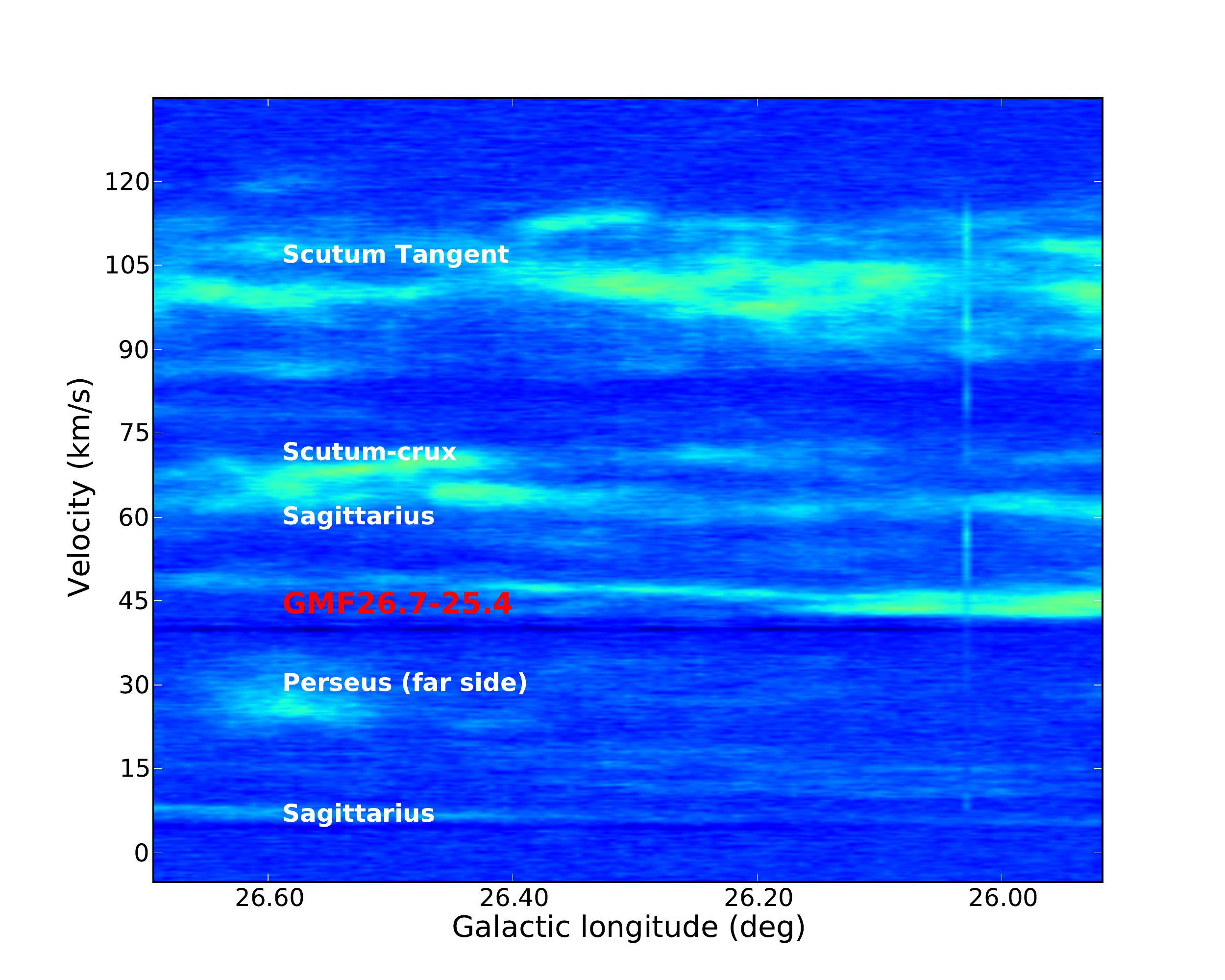}
\caption{\label{fig:pv} Position-velocity (PV) diagram (integrated over all latitudes) of the region shown in Figure~\ref{fig:higal} based on the Galactic Ring Survey $^{13}$CO(1-0) data. The approximate positions of the spiral arm features in this longitude range from the  \cite{Vallee2008} model are labeled in white, and the filament that we identify, GMF26.7-25.4, is labeled in red.}
\end{figure}

\section{Filament identification}

{\em Herschel} observations have recently highlighted the importance of filaments, as it tends to be the dominant morphology observed in star-forming clouds.  These filaments are comprised of cold gas and dust, the various signatures of which we will discuss below.  Defining a true filament, however, is not a trivial task and requires several steps to be confirmed. Below we take the example of the field near the F26.7-25.4 filament discovered in this work as an illustration.

Wide-field mid-infrared (MIR) images have proven to be a powerful tool in finding candidate structures like Nessie, which was identified in the 8\micron {\em Spitzer}/GLIMPSE image.  This phenomenon is illustrated in the top panel of Figure~\ref{fig:higal}, which shows the GLIMPSE 8\micron image of one of the regions near $l$ = 26\degr that we studied.  The contours are drawn in decreasing steps, highlighting the type of absorption structure that would be in catalogs using this method \citep[see][]{Simon2006, Butler2009, Peretto2009, Ragan2009}.  While this method has been successful in isolating the quiescent clouds in the Galactic plane, it alone may miss objects that could potentially have the same physical properties but lack the 8\micron background to be identified.  

Alternatively, extinction of starlight at near-infrared (NIR) wavelengths can be used to trace cold, intervening dust structures \citep[e.g.][]{Lombardi2001, Lombardi2009, Kainulainen2011b}.  In short, the amount by which starlight is dimmed is proportional to the column density of material along its line of sight.  This method is most reliable in the presence of many background stars, thus it also loses effectiveness with increasing Galactic latitude.

At longer wavelengths, the structures that were absorbing in the near and mid-infrared transition to optically thin emission. The {\em Herschel} HIGAL survey \citep{Molinari2010} 250\micron map of the same region is shown for comparison on the bottom panel of Figure~\ref{fig:higal}. While the same structure indeed does appear in emission here, so does a strong contribution from the warm dust associated with the active star formation region. In fact, because of the way dust emission depends on dust temperature, the emission from the warm dust tends to dominate the map, overwhelming the cold, compact emission that we aim to characterise. Therefore, in the following, we create our initial catalog of candidates using the NIR and MIR absorption methods.

Regardless of the candidate's initial identification, the true extent of a filament can only be judged after it is confirmed to be coherent in velocity space as well.  We selected the Galactic Ring Survey \citep[GRS][]{Jackson2006} which immediately limits the Galactic longitude range to 17.4\degr $\leq l \leq$ 55\degr.  We also utilised the \cite{Wienen2012} catalog of NH$_3$ observations of ATLASGAL \citep{Schuller2009} clumps and the \cite{Shirley2013} survey of Bolocam Galactic Plane Survey \citep[BGPS,][]{Rosolowsky2010,Aguirre2011,Ginsburg2013} clumps for a secondary confirmation of velocity coherence in higher-density gas tracers.  As it is not possible to {\it a priori} say to what degree the centroid velocity would change over the length of a genuine filament, instead of setting a maximum range, we require only that any velocity gradient be continuous to be considered ``coherent.''

Both the identification and verification steps outlined above are necessarily subjective, thus we can claim no statistical robustness or completeness.  The criteria we imposed were designed to identify 100\,pc scale quiescent structures.  In the following sections, we outline our method which was carried out by all co-authors by way of ``by-eye'' inspection of images.

\subsection{Creating the candidate catalog}

We first selected image databases that provide unbiased and contiguous coverage in the Galactic plane.  The {\em Spitzer} Galactic plane survey, GLIMPSE \citep{Benjamin2003}, is the ideal tool for identifying analogs to the known long filaments like Nessie \citep{Jackson2010}.  We used the GLIMPSE/MIPSGAL image viewer\footnote{\url{http://www.alienearths.org/glimpse}}. For a filament to be considered as a candidate, the absorption feature was required to extend $\sim$1\degr end to end and be identified by at least three group members. We allowed for gaps in extinction, usually at sites of star formation, if the extinction structure continued further.

As an alternative and supplement to the GLIMPSE data, we use the UKIDSS Galactic Plane Survey (GPS)\footnote{\url{http://surveys.roe.ac.uk:8080/wsa/gps_mosaic.jsp}} \citep{Dye2006, Warren2007a}, which provide large NIR image mosaics with large Galactic latitude range (up to $|b| \leq 2.5$\degr) in the northern hemisphere. The same $\sim$1\degr extent and three group member confirmation was required for filament candidates to be identified in this fashion.

As was the case for the serendipitous filament discovery by \citet{Tackenberg2013}, each filament contains a high-contrast extinction region, often previously studied as a relatively compact IRDC \citep[e.g.][in this case]{Beuther2002a}, but upon inspection of a wide-field mosaic, it was discovered to be part of a larger filamentary structure. The goal of this step was to explore the longest possible extent of the large-scale structure, which then requires confirmation as coherent velocity structures.
In total, our two methods yielded twelve candidates within the longitude range of the GRS.  They are listed in Table~\ref{tab:filament_candidates}. These candidates are certainly included in the catalogs compiled by \citet{Simon2006} and \citet{Peretto2009}. This paper is an attempt to identify long contiguous entities comprised of many smaller elements, such as was done by \citet{Tackenberg2013}. In the following section, we outline the method by which we test whether the candidates are single entities rather than multiple structures superposed along the sightline. 
 
\subsection{Filament coherence in velocity}

The second step of the process is to validate the velocity coherence using the GRS survey. The publicly-available GRS data\footnote{\url{http://www.bu.edu/galacticring/}} cover a velocity range from -5 to 85\,km s$^{-1}$ (or -5 to 135\,km s$^{-1}$ for $l <$ 40\degr).  We first created position-velocity (PV) diagrams over the full available velocity range within the longitude boundaries, integrating along the $b$ direction, an example of which is shown in Figure~\ref{fig:pv}. In some cases (e.g. GMF\,41.0-41.3) we constructed the PV diagram by integrating along the longitude direction. As is clear in Figure~\ref{fig:pv}, several possible features could correspond to the extinction structure. In order to determine which is the best match, we created integrated intensity maps for each of the velocity ranges, which we then plotted together with the extinction image.  Again, requiring verification from 3 group members, we selected the velocity features which corresponded best to the morphology of the feature seen 
in extinction or absorption.   

Figure~\ref{fig:pv} shows that features in a PV diagram can span different ranges in velocity, from a few to tens of km s$^{-1}$; we placed no restriction on the maximum width.  We list the full velocity ranges in Table~\ref{tab:coherent_filaments}, which can be as small as 5\,km s$^{-1}$ or as large as 13\,km s$^{-1}$.  Figures~\ref{fig:filament_16} through \ref{fig:filament_54} show the centroid velocity maps.  As with Nessie, in which the HNC $J$ = 1$\rightarrow$0 centroid velocities vary by less than 7\,km s$^{-1}$ over its $\sim$81\,pc length, we find similarly small overall gradients. In Section~\ref{s:results} we will discuss the properties of the final filament sample in more detail.

Through this step, we eliminated candidates F29.2-27.6, F25.9-21.9, F23.8-22.8, and F20.3-19.9 because they were not found to have a common velocity over the full length, but instead were superpositions of multiple velocity components.  The remaining eight candidates, two of which (F38.1-35.3 and F35.0-32.4) ended up as one velocity-coherent object, are henceforth termed ``Giant Molecular Filaments'' (GMFs) and are described in more detail in the next section. Parts of these verified GMFs correspond to catalog clouds \citep[e.g.][]{RomanDuval2009}. The aim of the present work is to identify the longest coherent structures which requires a different approach.

As a secondary verification, although we did not require it for GMF status, we also consult additional catalogs of dense-gas tracer observations, namely  the \citet[][hereafter W12]{Wienen2012} NH$_3$ survey of dense ATLASGAL clumps, Red MSX survey \citep[RMS][]{Lumsden2013}, or the \citet[][hereafter S13]{Shirley2013} N$_2$H$^+$ and HCO$^+$ follow-up survey of clumps in the Bolocam Galactic Plane Survey. Though these are catalogs of pointed observations rather than unbiased maps, these tracers more closely probe the densest structures, thus enabling us to validate the association of these emission peaks with the large scale coherent velocity structure seen in the GRS data.

\subsection{Biases in our filament identification}

Both extinction methods work best in the midplane of the Galaxy since the mid-infrared background and the density of background stars are both high.  Dense filaments that lie out of the Galactic midplane tend to lack the contrast to be identified in 8\micron extinction and may also lack a sufficient number of background sources to be appreciable in NIR extinction. Though we did not restrict the latitudes any narrower than the extent of the various surveys, the confirmed filaments have latitudes within a degree of 0\degr.  Another consequence of using the extinction method is that we are unlikely to detect structures any further than 6 to 8\,kpc \citep[see][]{Kainulainen2011b}.

Our candidate identification method also biases us in favour of identifying quiescent structures. Energetic star formation events can disrupt a cloud sufficiently such that it will no longer appear as an extinction structure.  For example, the G32.02+0.06 filament described in \cite{Battersby2012}, while in the longitude range of our search, was not identified as a candidate because it shows strong star formation activity, which disrupts the extinction signature that we sought in our hunt for quiescent structures.  

The GRS confirmation step also has its limitations.  Many sight lines in the Galactic plane exhibit multiple velocity components in $^{13}$CO, and with the inherent distance ambiguity problem \citep[e.g.][]{RomanDuval2009}, distance determination is far from straightforward. As mentioned above, we have adopted the near distance for the velocity component most closely matching the morphology of the absorbing structure.

\input{filaments_candidates_GRS_saga.tex}
\input{filaments_velo_coherent_saga.tex}

\section{Results}
\label{s:results}

\subsection{A sample of 7 velocity coherent filaments}

We find seven velocity coherent filaments in the longitude range of the GRS, the basic properties of which are listed in Table~\ref{tab:coherent_filaments}. In the Appendix, Figures~\ref{fig:filament_16} through \ref{fig:filament_54} show the GLIMPSE 8\micron image in greyscale.  We label sources associated with star formation, such as Westerhout objects \citep{Westerhout1958}, bright ($S_{12{\mu}m} > 100$\,Jy) IRAS sources, and \hii regions detected in the CORNISH survey \citep{Purcell2013}.  We show contours of $^{13}$CO integrated intensity (white) and ATLAGAL 870\micron emission (blue) to give an impression of the envelope and dense gas areas, respectively.

In the colour panels, the  first moment (centroid velocity) map of $^{13}$CO is displayed.  We plot the positions of all clumps from the Bolocam Galactic Plane Survey.  Using the \cite{Shirley2013} follow-up measurements of both the HCO$^+$ and N$_2$H$^+$ tracers, the BGPS clumps are marked with circles if their adopted $v_\mathrm{lsr}$ lies within the velocity range of the GMF, and they are marked with an $\times$ if the velocity is outside of that range.  We mark with diamonds the positions of W12 NH$_3$ clumps with velocities in the range. This secondary check against the catalogs of dense gas tracers helps us to validate that the dense gas is at the same distance as the $^{13}$CO cloud from the GRS. 

In Figures~\ref{fig:fil16_pv} through \ref{fig:fil54_pv}, we show the position velocity diagrams for each GMF for the full velocity range that we inspected with the GRS $^{13}$CO data. We highlight the velocity range that is associated with the GMF morphology. 

To place the filaments in a Galactic context, we display latitude of the physical Galactic midplane when possible.  Because the Sun lies $\sim$25\,pc above the Galactic midplane, the Galactic coordinate ($l$,$b$) = (0,0) does not actually coincide with the location of the Galactic centre \citep{Blaauw1960}.  Consequently, at the typical distances of the structures in this sample, the physical midplane is closer to $b \sim$ 0.5\degr.  In most cases, the filaments we identify lie above the real midplane.  We note the average projected height off of the plane in parsecs in Table~\ref{tab:galactic}. 

Finally, we note the association in longitude and velocity space to the predicted positions of the spiral arms of the Milky Way as compiled by \cite{Vallee2008}.  In most cases, we find little or no association between the filaments and spiral arm structures, in contrast with the recent findings of \cite{Goodman2013} regarding the Nessie filament. In the following, we describe each GMF individually, noting their association (or lack thereof) with other well-known regions of star formation and the predicted positions of spiral arms.

\subsubsection{GMF\,18.0-16.8}

This filament is at the edge of the longitude coverage of the GRS, but the supplemental velocity information provided by the W12 catalog confirms that it extends to $l \sim$ 16.8\degr. Figure~\ref{fig:filament_16} shows this filament in a narrow velocity range (21-25 km s$^{-1}$). To the north of this filament lies M16, also known as the Eagle Nebula, estimated to be roughly 2\,kpc from the Sun \citep{Hill2012}, which is in agreement with the distance derived to the rest of the structure we find.  We find 16 BGPS clumps from the S13 catalog and 6 clumps from the W12 catalog that have consistent velocities. 

This structure lies well above the location of the Galactic midplane.  According to the \cite{Vallee2008} model, the velocity of this filament appears to agree with the predicted value of the Perseus spiral arm, however at this longitude the Perseus arm begins from the ``far'' side of the Galaxy at 12\,kpc distance. Due to its close association with M16 and strong extinction features, we maintain our adoption of the ``near'' kinematic distance, which means it is unrelated to the Perseus arm. 

\subsubsection{GMF\,20.0-17.9}

GMF\,20.0-17.9, shown in Figure~\ref{fig:filament_17}, was already identified in part by \cite{Tackenberg2013} and exhibits an arc-like projection between two ``bubble'' structures.  There are 39 BGPS clumps within the velocity range, and 16 ATLASGAL clumps exhibiting consistent velocities in NH$_{3}$ (W12).  This field also contains 100 BGPS clumps with large velocity offsets from the given range.  In fact, each MIR-bright region has associated \hii regions with velocities around 65\,km s$^{-1}$ \citep{Urquhart2011}, and thus appear to be unrelated to the filament we find at 37 to 50\,km\,s$^{-1}$. W39, on the other hand, has a consistent velocity \citep{Purcell2013}. We note that there is an indication of a velocity jump for a sector of this GMF near $b \sim -0.2$\degr, 19.2\degr $< l <$ 19.9\degr. If one instead assumes that this region is not associated with the rest of the GMF, the total ATLASGAL emission drops by 20\%, and the total $^{13}$CO emission decreases by 12\%, thus their ratio degreases by 10\

The Galactic midplane appears to intersect with this filament, and its velocity agrees fairly well with that of the Scutum-Centaurus (SC) spiral arm \citep{Vallee2008}.  It is possible, like in \cite{Li2013}, that a massive star formation event, in this case probably associated with W39, may have carved the arc-like structure seen in this filament. 
 
\subsubsection{GMF\,26.7-25.4}

While GMF\,26.7-25.4, shown in Figure~\ref{fig:filament_26} appears parallel to the Galactic plane, it lies $\sim$1.3\degr, or $\sim$68\,pc, above the physical midplane.  Also as a consequence of its high latitude, it is poorly covered by the BGPS spectroscopic survey, and the W12 catalog only finds one source with a consistent velocity. This also appears to be one of the most quiescent filaments in that there are no prominent IR sources, except IRAS\,18348-0526,  and the only two \hii regions cataloged in the CORNISH survey appear to be at different velocities according to the corresponding W12 measurements. 

\subsubsection{GMF\,38.1-32.4: two overlapping structures}

Perhaps unsurprisingly, at longitudes near the meeting place of the bar and spiral arm of the Galaxy, this region has multiple objects of interest. Between longitudes 32\degr and 38\degr, three candidates were identified in extinction (see Table~\ref{tab:filament_candidates}).  Upon examination in the GRS data, we found that F38.1-35.3 and F35.0-32.4 were connected in velocity space (between 50 and 60\,km\,s$^{-1}$), and F35.3-34.3, which overlaps with the previous two, had distinct velocities offset to lower values (43 to 46\,km\,s$^{-1}$).  While the latter structure (GMF\,38.1-32.4b) is not especially elongated along the Galactic plane, it still has length $\sim$80\,pc.  The former (GMF\,38.1-32.4a), which is very much extended along the midplane, is the longest and most massive object in our sample.  We show both velocity components in Figure~\ref{fig:filament_38}. 

The different velocity ranges indicate that these filaments lie at different distances. The larger filament (GMF\,38.1-32.4a) is also further away and has 85 BGPS (S13) clumps and 12 W12 clumps within its velocity range.  W44 is a massive dense clump at 3.7\,kpc \citep{Solomon1987}, consistent with the distance of GMF\,38.1-32.4a. The smaller filament (GMF\,38.1-32.14b) coincides with 9 BGPS clumps in its velocity range. Since it is nearer than the first, the projection of the Galactic midplane (dashed line) is at slightly lower latitude with which it intersects. There are another 149 BGPS clumps whose velocities coincide with neither of the filaments identified here, and there are 18 \hii regions detected by CORNISH, but given the confusion in this region, it is difficult to determine which are associated with the filaments. 

\subsubsection{GMF\,41.0-41.3}

This filament extends mainly perpendicular to the Galactic plane, unlike most of the rest of the sample. There are three BGPS sources with consistent velocities, one W12 NH$_3$ clump, but no CORNISH \hii regions. As shown in Figure~\ref{fig:filament_41}, the projected Galactic midplane just barely intersects with the southernmost tip of the filament. 

The central IR-bright region corresponds to the position of supernova remnant 3C397 and the BGPS clumps in that region are at velocity of $\sim$60\,km s$^{-1}$, well outside of the selected velocity range for the filament. This filament corresponds to molecular gas that is thought to be in the foreground of the remnant, which also serves to confine the remnant \citep{Jiang2010}. Otherwise, this filament appears rather quiescent. 

\subsubsection{GMF\,54.0-52.0}

Part of this filament has been identified recently by \cite{Kim2013ppvi}, who find it is in an early phase of cluster formation relative to standard local molecular clouds. GMF54.0-52.0 shares a common projected area with the 500\,pc gas wisp identified in \cite{Li2013}, which coincides with the Perseus spiral arm on the far side of the Galaxy \citep{Dame2001}. Here, we study gas in a different velocity range (20 to 26\,km\,s$^{-1}$) displayed in Figure~\ref{fig:filament_54}.

Within this structure, there are 18 BGPS clumps which have velocities between 20 to 26\,km s$^{-1}$. Three \hii regions were detected in this range in the CORNISH survey, but none appear to have consistent velocities. W52 is projected in the vicinity, but there is not independent measurement of its velocity in the literature. The IR-bright region to the west resides at a different velocity (2-6\,km s$^{-1}$) and thus appears to be a unassociated region of (high-mass) star formation \citep{Urquhart2011}. Therefore, although the IR image would lead one to conclude that this filament is vigorously forming stars, it appears largely devoid of the signposts explored here.

\subsection{Physical properties}
\label{ss:props}

Next we derive the basic physical properties of each GMF.  By first deriving the kinematic distance, we can then calculate the size, mass and density of each GMF.  We provide the values for the analogous structures Nessie \citep{Jackson2010} and G32.02+0.06 \citep{Battersby2012} for comparison.

\subsubsection{Kinematic distance, length, and mean velocity gradient}

Our sample of filaments exhibits large-scale velocity coherence in that they have relatively smooth velocity gradients in a continuous elongated structure. We use the velocities to derive a kinematic distance range according to the BeSSeL results \citep{Reid2009} assuming the standard Galactic parameters. The velocity ranges and the corresponding kinematic distance ranges are listed in Table~\ref{tab:coherent_filaments}.

We measure the angular length from end to end of the significant $^{13}$CO emission (above 1\,K\,km\,s$^{-1}$) in the selected velocity range and convert these to physical length using the mean kinematic ``near'' distance.  We have not attempted to correct for projection effects, thus the lengths reported in Table~\ref{tab:coherent_filaments} are lower-limits. The maximum projected length is 232\,pc for GMF\,38.1-32.4a, which interestingly is one of the filaments that agrees most closely to the location of the Galactic midplane. Determining the width of the GMFs, which is substantially different between the dense gas structure and the large scales traced by $^{13}$CO, is not a trivial task and beyond the scope of this paper. A comprehensive study of the widths, their relationship with each other and to their galactic environment, will be addressed in a future paper.

Most filaments exhibit a velocity gradient along their length. Examining Figures~\ref{fig:fil16_pv} through \ref{fig:fil54_pv}, we find that GMF\,18.0-16.8 and 38.1-32.4a show decreasing velocities with increasing longitude, while GMF\,20.0-17.9, 26.7-25.4, 38.1-32.4b, and 54.0-52.0 show increasing velocities with increasing longitude. GMF\,41.0-41.3 shows no trend either way. We compute the average velocity gradient over the entire length of the filament, which are reported in Table~\ref{tab:coherent_filaments}.  We find values between 40 and 80\,km\,s$^{-1}$\,kpc$^{-1}$, and longer filaments tend to exhibit larger mean velocity gradients. No correlation exists between velocity gradient and longitude or distance.  We list the values in Table~\ref{tab:coherent_filaments}.

\input{filaments_galactic_saga.tex}

\subsubsection{Mass}

We calculate the filament mass in two ways: first using dust emission, which is a useful probe the densest gas, and second using the $^{13}$CO emission to probe the total mass in the cloud. To restrict our measurements to equivalent areas, we first create masks using the significant $^{13}$CO emission ($>$ 1\,K km s$^{-1}$) in the indicated velocity ranges (see Table~\ref{tab:coherent_filaments}), which is the area shown in colour in the centroid velocity panels in Figures~\ref{fig:filament_16} to \ref{fig:filament_54}. 

In order to determine the total amount of {\it dense} gas, we use dust emission from the ATLASGAL 870\micron dust emission survey \citep{Schuller2009}. Due to filtering out of large-scale emission, the ATLASGAL data are sensitive primarily to the densest gas.  After the $^{13}$CO masking, we uniformly impose an additional emission threshold of 250\,mJy (5\,$\sigma$) and compute the column density assuming a temperature of 20\,K, gas-to-dust ratio of 100, and a dust opacity at 870\micron of 1.42\,g\,cm$^{-2}$, interpolated from \citet{Ossenkopf1994}.  With these assumptions, our column density sensitivity is of the order 10$^{22}$\,cm$^{-2}$.  Assuming the mean distance to each filament, we calculate the total dense gas mass.\footnote{Note that in the case of F38.1-32.4, since there are two overlapping filaments at different velocity ranges, the computed dense gas mass for each structure includes a  contribution from the unrelated cloud.}  We find dense gas mass values that range between 8 and 500 $\times$ 
10$^2$\msun. They are listed in Table~\ref{tab:galactic}. 

To calculate the mass of the cloud, we use the $^{13}$CO emission above 1\,K km s$^{-1}$ within the selected velocity range.  $^{13}$CO(1-0) is considerably more widespread than the dust emission because it is excited in low density ($n <$ 10$^3$\,cm$^{-3}$) gas.  To first find the column density of $^{13}$CO, we follow \citet{RohlfsWilson2004} formulation. We use the $T_\mathrm{ex}$ values derived in \citet{RomanDuval2010}, which assumes $T_\mathrm{ex}$($^{12}$CO) = $T_\mathrm{ex}$($^{13}$CO), for the corresponding clouds in the GRS catalog that match the position and velocity ranges of the GMFs.  We adopt the $^{12}$C/$^{13}$C ratio as it varies with Galactocentric radius ($R_\mathrm{gal}$, given in Table~\ref{tab:galactic}) from equation 3 in \citet{Milam2005}\footnote{The observed gradient with Galactocentric radius was fit with he following linear relation $^{12}$C/$^{13}$C = 5.41\,$R_\mathrm{gal}$\,+\,19.03.}, which gives values between 46 and 59.  We assume a uniform ratio of $n$($^{12}$CO)/$n$(H$_2$) 
= 1.1 $\times$ 10$^{-4}$ \citep{Pineda2010}. Together, these assumptions result in an H$_2$ column density sensitivity of $\sim$10$^{20}$\,cm$^{-2}$, then we compute the total (H$_2$) cloud masses using the GMF's mean kinematic distance.  The masses, which range between 4.9 $\times$ 10$^4$ and 7 $\times$ 10$^5$\msun, are listed in Table~\ref{tab:galactic}.

\subsubsection{Dense gas mass fraction (DGMF)}

A number of recent studies of star formation in Milky Way molecular clouds have arrived at the consensus that one key quantity governing the efficiency with which a cloud forms stars is the fraction of the ``dense'' gas in a molecular cloud \citep{Goldsmith2008, Heiderman2010, Lada2010, Lada2012}, often referred to as the dense gas mass fraction (DGMF).  We calculate the DGMF by taking the ratio of dense gas to total cloud mass over the same areas. We find DGMF values between 2.9 and 18.5\%, with a mean of 5.4\%. These values are generally consistent with the DGMF found in local star-forming clouds \citep{Kainulainen2009, Lada2010}, high-mass star-forming clumps \citep{Johnston2009, Battisti2014}, and other large molecular filaments \citep{Kainulainen2011b, Battersby2012}.  

\begin{figure}
\includegraphics[width=0.5\textwidth]{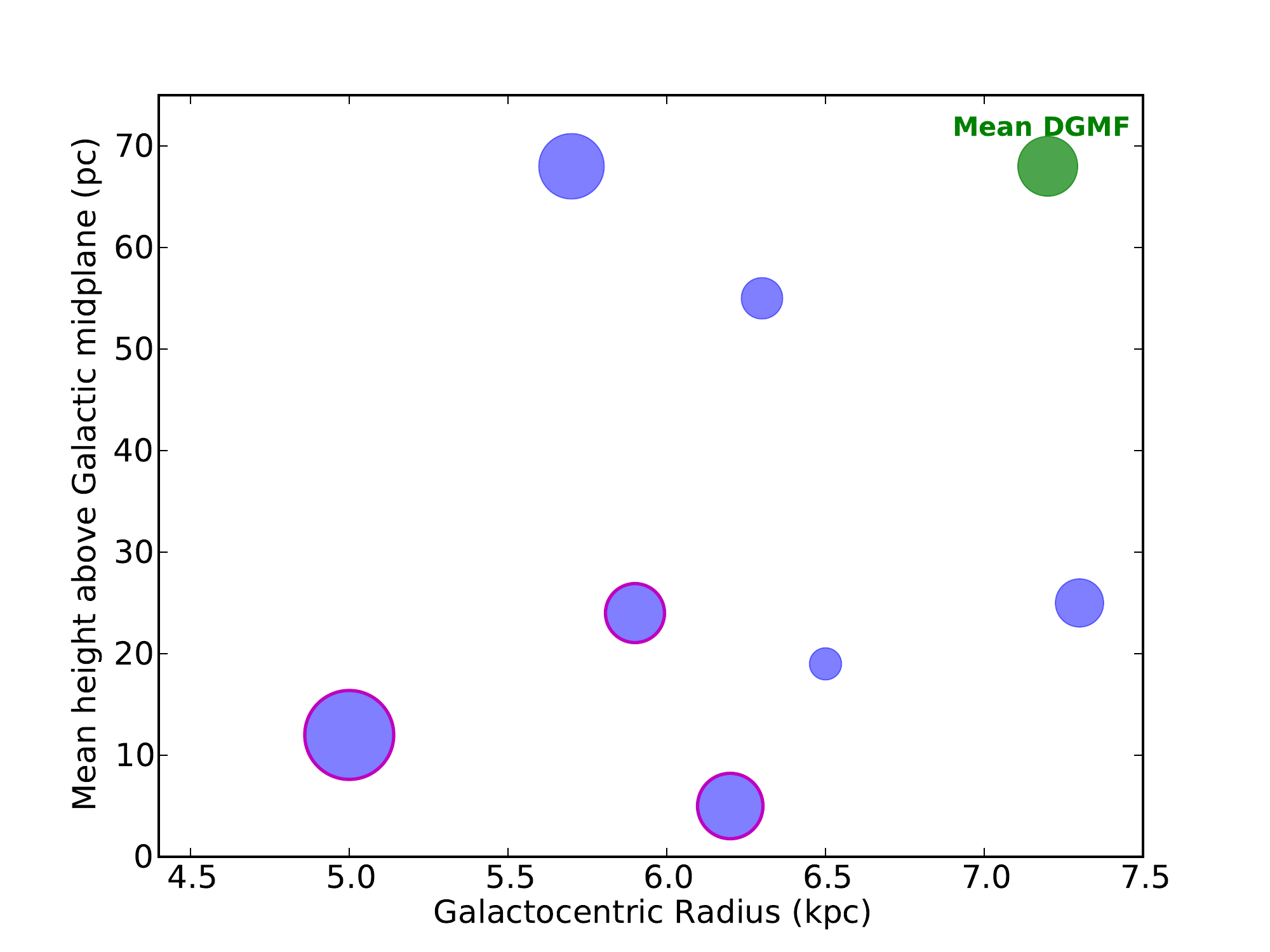}
\caption{\label{fig:dgmf_rp} The mean height of the GMF above the physical midplane versus its mean Galactocentric distance. The area of the marker is proportional to the DGMF, where the mean value (5.4\%) is shown in the upper right corner. The points circled in red are the GMFs that intersect with the midplane (shown in parentheses in Table~\ref{tab:galactic}).}
\end{figure}

While the DGMF tends to go down with increasing Galactocentric radius, we also find that GMFs centred (in latitude) closer to the Galactic midplane tend to have higher DGMF values than those significantly out of the plane. These trends are visualised in Figure~\ref{fig:dgmf_rp}, in which we show the height above the plane as a function of Galactocentric radius. The size of the circle corresponds to the DGMF compared to the mean (shown in the upper right). The most extreme example of a filament out of the plane, the perpendicular filament GMF\,41.0-41.3, has the smallest DGMF (1.6\%) while GMF\,20.0-17.9, the filament closely associated with the midplane (and also the SC spiral arm, more on this in Section~\ref{ss:spurs}) has a DGMF of 12.0\%.  Nessie -- a filament found exactly within the SC arm in the fourth quadrant -- exhibits a DGMF of 50\% \citep{Goodman2013}, though because HCN emission was used to compute the total cloud mass (instead of $^{13}$CO emission as in this paper) this value may be inflated. 
 Still, it appears that the DGMF is connected to the location and environment of the filament. We acknowledge that these trends are weak and statistically robust with the present data, thus we refrain from any further quantitative analysis.

\begin{figure}
\includegraphics[width=0.5\textwidth]{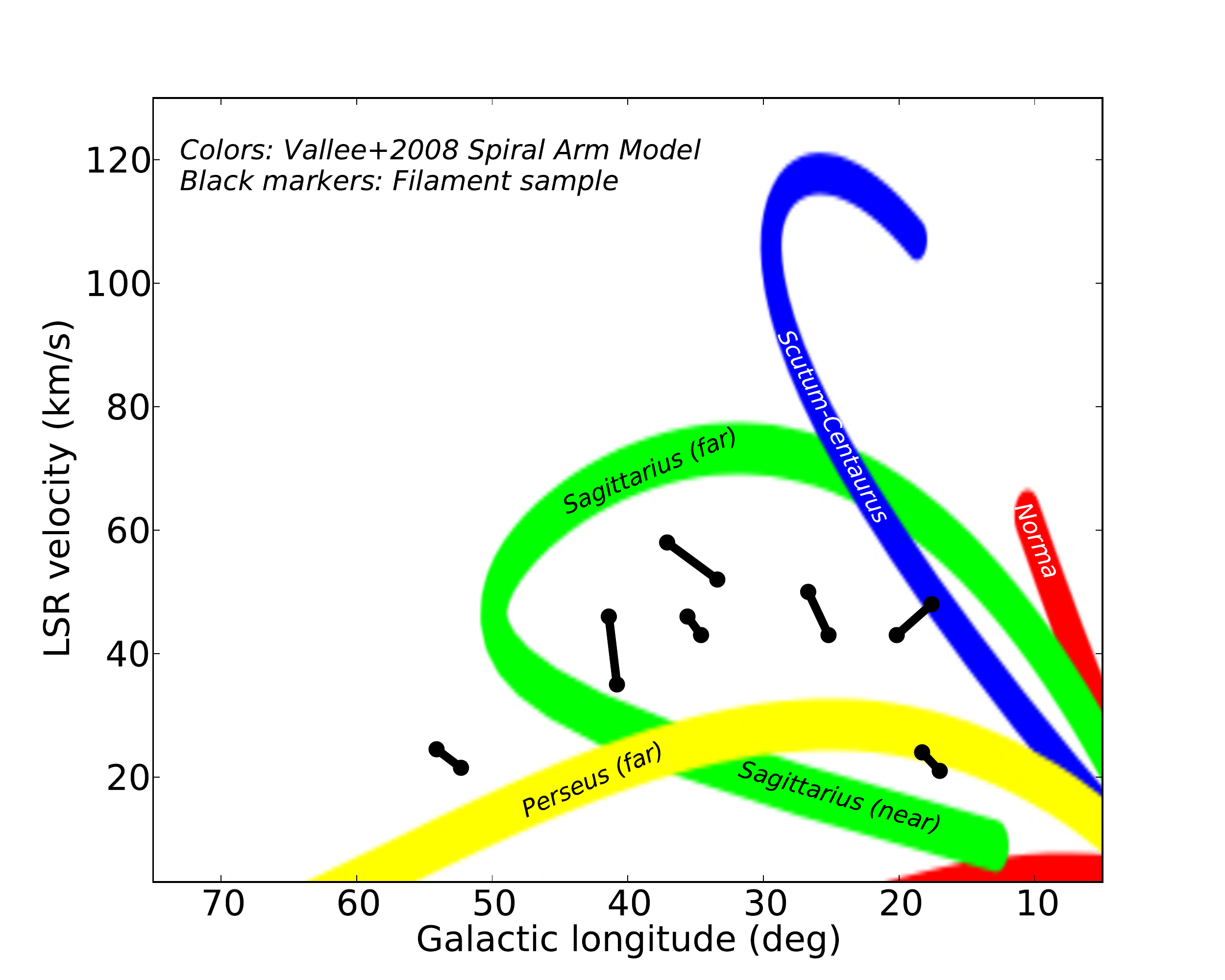}
\caption{\label{fig:vallee} Illustration of the predicted LSR velocities of the Norma (red), Scutum-Centaurus (blue), Sagittarius-Carina (green), and (far) Perseus (yellow) spiral arms as a function of Galactic longitude in the first quadrant taken from \citet{Vallee2008}. Each set of two black circles represent the filament sample, taking approximate values of $v_\mathrm{lsr}$ from the ends of the filaments.}
\end{figure}

\section{Discussion}

\subsection{Filaments in the Galactic context}
\label{ss:galactic}

We investigate how the GMFs fit into the Milky Way structure.  In Figure~\ref{fig:vallee} we plot the $v_\mathrm{lsr}$ range for each filament as a function of the Galactic longitude. For reference, we show the \citet{Vallee2008} predicted velocity-longitude loci associated with each of the four spiral arm structures that pass through the first quadrant. We attempt to represent the orientation of the GMFs by estimating the mean velocity on both ends of the filament, but the velocity maps are clumpy and quite irregular, so we stress that this aspect is very approximate. Note that in this quadrant, the Perseus arm (yellow) indicates the component on the far side of the galaxy.  We find that the GMFs do not correlate well with the spiral arm loci (see also Figure~\ref{fig:pv}).  One exception, GMF\,20.0-17.9, appears to intersect with the locus of the SC arm at low-$l$ end. If one does the same exercise for Nessie in the fourth quadrant of the Galaxy, the filament corresponds precisely to the SC arm prediction, 
as asserted by \citet{Goodman2013}.

The schematic shown in Figure~\ref{fig:vallee} shows where the arm has its peak density, but the actual arm width and its projection into $v_{lsr}$ are not well-defined.  It is thus worth exploring the margin by which the GMFs seem to deviate from the arms.  Estimates of spiral arm width in the literature range from 0.1 to 0.4\,kpc \citep{Vallee2008, Reid2009}.  Typical uncertainties in the kinematic distance method are of order 0.5\,kpc, so up to a 1\,kpc range uncertainty must be allowed for in this discussion.  GMF\,18.0-16.8, GMF\,20.0-17.9 and GMF\,26.7-25.4 are in the longitude range intersecting with the SC arm, and at those longitudes, the SC arm should roughly lie 3.5, 3.7 and 4.3\,kpc from the Sun. The median distances to those GMFs are 2.3, 3.5 and 3.1\,kpc, so only GMF\,20.0-17.9 falls within the uncertainty range. Similarly the nearer Sagittarius arm lies at 1\,kpc at $l$=18\degr, gradually increasing to 2\,kpc at $l$=41\degr, yet the GMFs are all more distant than this arm by more than 1\,kpc (
except GMF\,41.0-41.3 which is $\sim$700\,pc from the Sagittarius arm locus). Even with these considerations, the GMFs appear to be largely inter-arm in nature: more distant the nearby Sagittarius spiral arm and foreground to the prominent SC arm.

We find that most filaments are centred near or above the physical Galactic midplane.  The approximate position of the physical midplane is shown in Figures~\ref{fig:filament_17}, \ref{fig:filament_38}, and \ref{fig:filament_41}; in the other cases, the midplane is too far south to be shown in the panel.  In Table~\ref{tab:galactic}, we compute the average height above the midplane in parsecs.  There are three cases -- GMF\,20.0-17.9 and GMF\,38.1-32.4 (a and b) -- where the filament intersects with the estimated position of the midplane of the Milky Way. GMF\,20.0-17.9 and GMF38.1-32.4a are also the two most massive and two longest filaments in the sample.  Otherwise, the filaments do not agree well with the Galactic midplane and, as in the case of GMF\,41.0-41.3, can be oriented more perpendicular to the plane. If one performs the same exercise with Nessie, it exactly coincides with the latitude range of the Galactic midplane \citep{Goodman2013}.

That we have identified most of the GMFs are above the physical midplane is perhaps not surprising given the parameters of most Galactic plane surveys.  It is unlikely that GMFs preferentially reside above the plane.  Due to the fact that most Galactic Plane Surveys to date have a narrow latitude coverage centred on $b=0$\degr while the true midplane is in fact at negative latitudes due to the Sun's height above the plane, analogous structures below the midplane would reside at latitudes $b < 0.5$\degr.  

Though our statistics are limited, the masses and DGMFs for GMFs within the physical midplane tend to be higher than for those out of the plane (GMF\,26.7-25.4 seems to be the exception here, with a high DGMF and large offset from the plane).  What seems to play an even stronger role (though our statistics are limited) is the association with spiral arms. For instance, GMF\,20.0-17.9 is the single filament we find that is near a spiral arm, both in longitude-velocity space and latitude. It also has the highest DGMF of the sample.  Nessie, another bona-fide spiral arm filament \citep{Goodman2013} has a DGMF of roughly 50\%.  The tendency of spiral arm clouds to have higher DGMFs would be consistent with what is found in M51 by \citet{Hughes2013}. They found that molecular clouds within the spiral arms have a higher fraction of gas at high densities than inter-arm clouds.  If this trend for GMFs to have higher DGMF within spiral arms is validated with a larger sample, it would indicate that the position of a (
filamentary) cloud structure with respect to the spiral arms plays a central role in determining its DGMF and hence its star formation.

\subsection{GMFs: Milky Way inter-arm clouds or spurs?}
\label{ss:spurs}

One of the fundamental questions in studies of spiral galaxies is what role (if any) do the spiral arms play in the production of stars in galaxies? While molecular gas is observed to be more concentrated in the spiral arms, whether the arm itself plays a direct role in inducing star formation is not clear. The inter-arm nature of the GMFs we have identified make them important population with which to investigate the role of spiral arms in the Milky Way, as they are probably more appropriate clouds to compare with those studied in external galaxies than nearby molecular clouds. For example, the PAWS survey of M51 \citep{Schinnerer2013} represents the current state-of-the-art in extragalactic studies, reaching 40\,pc resolution and a mass sensitivity of 1.2\,$\times$\,10$^5$\,M$_{\odot}$ \citep{Colombo2014a}. If the GMFs resided in M51, most would be resolved and detectable.

Inter-arm cloud populations have been scrutinised observationally in external galaxies \citep{Elmegreen1980, Aalto1999, Scoville2001, Shetty2007, Foyle2010, Hughes2013, Colombo2014a}, in the Milky Way \citep{RomanDuval2010, Moore2012, Eden2013} and in numerical simulations \citep[e.g.][]{KimOstriker2002, Chakrabarti2003, Dobbs2006, Shetty2006, Smith2014}. The history of inter-arm clouds is related to so called ``spurs'' or ``feathers'' emanating from the spiral arms, which have been credited to either the growth of gravitational or magneto-Jeans instabilities preferentially perpendicular to the arm \citep{Balbus1988} or alternatively, to the rotational shearing of over-densities in the spiral arm itself \citep{Dobbs2006, Shetty2006}.  While the exact properties and lifetimes of spurs and inter-arm features are subject to ongoing theoretical work, only recently have observations been able to systematically study inter-arm clouds in external galaxies at high enough angular resolution to probe individual 
molecular cloud scales \citep[e.g. the PAWS survey,][]{Schinnerer2013}.  

Both observations and simulations of spiral galaxies indicate that spurs and strong inter-arm features are most prominent on the trailing side of spiral arms in the outer regions of disk galaxies.  The situation for inter-arm clouds appears to get significantly more complex in the inner regions of galaxies, where spiral arms are more tightly wound.  For an illustrative example, one need not look further than the two-arm Whirlpool galaxy, M51, whose structure shows high complexity in the inner few kiloparsecs \citep{Aalto1999, Schinnerer2013}.  Since the spiral arms are closer to one another and the kinematics are more complex in the inner regions, assigning inter-arm clouds to one particular arm is much more difficult.  Further investigations are needed into the projected kinematic signatures of spur or feather formation in the Milky Way's plane in order to connect inter-arm clouds to clouds within the arms.

GMF\,20.0-17.9 presents the most compelling case of a ``true spur'' in our sample, as it is the only GMF that (still) may have a physical connection to a spiral arm. It exhibits a velocity range (37 - 50 km s$^{-1}$) that is close to that of the SC arm (see Figure~\ref{fig:vallee}) and is also near the midplane of the the Galaxy.  However, the sense of the velocity gradient goes against the trend expected if the GMF were a component of the arm itself (like Nessie).  Figure~\ref{fig:vallee} shows that the $l$ $\sim$ 18.0\degr end aligns with the SC arm's predicted position, but rather than continuing to higher values with the arm, the velocities in GMF\,20.0-17.9 decrease with increasing longitude.  In addition, the arm-end of this GMF corresponds to the position of a massive GRS cloud G18.04-0.36 identified in \citet{RomanDuval2009, RomanDuval2010}, which lies in the SC arm.  This is an enticing clue that may helpful in studying different spur-formation scenarios, particularly mechanisms which shear out gas 
streams from massive molecular clouds within the arms \citep{Shetty2006}, but we need much better statistics to test these scenarios.

The arguments provided above that have led us to conclude that GMFs are inter-arm in nature at the same time lead us to also conclude that Nessie is a spiral arm filament.  Nessie has the correct distance, latitude, and velocity gradient (in agreement with sense of the arm) to be a part of the SC spiral arm, in agreement with the findings of \citet{Goodman2013}.  Because we only know of one long filament in the fourth quadrant and a similar study as presented here has not yet been performed on the fourth quadrant, we can only speculate as to why we found no Nessie-like clouds in the first quadrant despite the similarity in discovery methods.  Are we just not as sensitive to spiral arm filaments in the first quadrant compared to the fourth? It is possible that the frequency and orientation of spiral arm filaments and spurs is different in the two quadrants. In the first quadrant, the SC arm is wound tighter and thus closer to the Galactic centre. It could be that inter-arm clouds are more frequent or 
prominent in this part of the Galaxy such that they are the filaments that one is more likely to find in absorption.  Our orientation with respect to the SC arm changes in the fourth quadrant, which may effect our ability to find inter-arm GMFs, so it will be interesting to extend this method to this sector of the Galaxy. 


\section{Conclusion}

Filamentary clouds appear to be intimately tied to star formation, but to date we have not had an observational consensus of how long and massive the filaments in the Milky Way can be.  Are they simply larger versions of the smaller filamentary molecular clouds in local star-forming regions, or does their role as the building blocks of the Galaxy shape their properties and determine their fate?  In order to understand the part they play in star formation in the Milky Way, many such objects must be studied in detail.  This paper presents the first catalog of filaments identified as extinction signatures using the {\em Spitzer}/GLIMPSE and UKIDSS Galactic plane surveys and verified in their coherent velocity structure using the Galactic Ring Survey.  A similar approach can be taken with complementary surveys in other interior parts of the Galactic plane.  

We introduce a sample of Giant Molecular Filaments (GMFs) based on initial identification in mid-infrared extinction seen in the {\em Spitzer}/GLIMPSE images and confirmation of velocity coherence using $^{13}$CO data from the Galactic Ring Survey. We require an angular length of at least 1 degree, and after confirmation we found that these structures ranged from 60 to 230\,pc in length. We calculate their total masses in using both the $^{13}$CO and complementary dust emission data at 870\micron from the ATLASGAL survey to trace the total cloud and the dense gas mass, respectively. We find the ratio between the two (the DGMF) to range between 2 and 12\%, consistent with measurements in local star-forming clouds and a recent estimate of the galactic mean \citep{Battisti2014}.

Most filaments have some association with -- at most -- a low level of star formation. As our technique requires the GMFs to appear in extinction, we are especially sensitive to quiescent clouds.  Using their positions and velocities, we place the GMFs in the Galactic context using a model of the Milky Way's kinematic structure.  We find that most GMFs appear to be spiral arm spurs or inter-arm clouds to which extinction studies may be especially sensitive in the first quadrant due to our orientation with respect to the Scutum-Centaurus spiral arm.  The DGMF in the GMFs is tentatively correlated with their environment: the closer a GMF is to the physical Galactic midplane and to a spiral arm structure, the higher its DGMF tends to be, though better statistics are needed to confirm this in the entire Milky Way plane.  If this trend is genuine, it would mirror what is observed in M51 by \citet{Hughes2013} where the molecular clouds within spiral arms have a higher fraction of dense gas than inter-arm clouds.  
As such, the GMFs could play an important role in connecting the Milky Way to other galaxies and provide a tool for studying the small scale effects of feedback and dynamics in the inter-arm regions.  This sample of seven GMFs is a starting point in thoroughly mining the Galactic plane in all quadrants, which will allow for a statistically robust study of the link between GMFs and Galactic structure.

\begin{acknowledgements}
The authors thank Rahul Shetty, Fabian Heitsch, Rowan J. Smith, and Clare Dobbs for useful discussions. SR and JK are supported by the Deutsche Forschungsgemeinschaft priority program 1573 (``Physics of the Interstellar Medium''). This publication makes use of molecular line data from the Boston University-FCRAO Galactic Ring Survey (GRS). The GRS is a joint project of Boston University and Five College Radio Astronomy Observatory, funded by the National Science Foundation under grants AST-9800334, AST-0098562, \& AST-0100793.  This paper also made use of information from the Red MSX Source survey database at www.ast.leeds.ac.uk/RMS which was constructed with support from the Science and Technology Facilities Council of the UK. This research made use of APLpy, an open-source plotting package for Python hosted at http://aplpy.github.com. This research has made use of the SIMBAD database, operated at CDS, Strasbourg, France. 
\end{acknowledgements}


\input{filaments_saga.bbl}
\appendix

\section{Image gallery}
\input{filaments_images_saga.tex}

\input{filaments_pv_saga.tex}
\end{document}

%% file: filaments_candidates_GRS_saga.tex
\begin{table}[htbp]
  \centering
  \caption{Extinction filament candidates within GRS longitude range \label{tab:filament_candidates}} 
  \begin{tabular}{c c c c c c }
    \hline
    Initial name& Lower $l$ & Upper $l$ & Lower $b$ & Upper $b$ & \\
    & [ \degr ] & [ \degr ] & [ \degr ] & [ \degr ] & \\
    \hline
    F18.0-17.5 &  17.4 &  18.3 &  +0.0 &   +0.7  & \\    
    F20.3-19.9 &  19.5 &  20.4 &  -1.3 &   +0.2  & \\    
    F20.0-17.9 &  17.6 &  20.2 &  -0.7 &   +0.3  & \\    
    F23.8-22.8 &  22.1 &  23.8 &  +0.8 &   +2.0  & \\    
    F25.9-21.9 &  21.7 &  26.2 &  -1.0 &   +0.5  & \\    
    F26.7-25.4 &  25.2 &  27.0 &  +0.5 &   +2.2  & \\    
    F29.2-27.6 &  27.2 &  29.4 &  -0.5 &   +0.5  & \\
    F35.0-32.4 &  32.3 &  36.0 &  -0.3 &   +0.8  & \\
    F35.3-34.3 &  33.7 &  35.6 &  -2.0 &   +0.4  & \\  
    F38.1-35.3 &  34.6 &  38.9 &  -0.9 &   +0.6  & \\
    F41.0-41.3 &  40.8 &  41.4 &  -0.8 &   +0.5  & \\
    F54.0-52.0 &  48.0 &  54.4 &  -0.5 &   +0.7  & \\
\hline
  \end{tabular}\\
  \footnotesize{Column 1: Name given to filaments based on initial
    identification; Column 2-5: Galactic coordinates of the boundaries of the  
    extinction-identified filaments.}
\end{table}

%% file: filaments_velo_coherent_saga.tex
\begin{table*}[htbp]
  \caption{GMFs with coherent velocities \label{tab:coherent_filaments}} 
\begin{center} 
 \begin{tabular}{lccccccccc}
    \hline
    Initial name & 
    Lower $l$ & 
    Upper $l$ & 
    Lower $b$ & 
    Upper $b$ & 
    Angular length & 
    Velo. range & 
    Distance &
    length &
    $< \bigtriangledown v >$ \\
    & [ \degr ] & [ \degr ] & [ \degr ] & [ \degr ] & [ \degr ] 
    & [km s$^{-1}$] & [kpc] & [pc] & [km s$^{-1}$ kpc$^{-1}$] \\
    \hline
    GMF18.0-16.8  &  16.4 &  18.3 &   0.0 &   +1.2  & 2.5  & 21 - 25 & 2.1 - 2.4 & 88 & 45  \\
    GMF20.0-17.9  &  17.6 &  20.2 &  -0.7 &   +0.3  & 1.8  & 37 - 50 & 3.3 - 3.7 & 170 & 76 \\
    GMF26.7-25.4  &  25.2 &  26.7 &  +0.5 &   +2.2$^1$ & 2.0  & 41 - 51 & 2.9 - 3.3 & 123 & 82 \\
    GMF38.1-32.4a &  33.4 &  37.1 &  -0.4 &   +0.6  & 3.8  & 50 - 60 & 3.3 - 3.7 & 234 & 43 \\
    GMF38.1-32.4b &  34.6 &  35.6 &  -1.0 &   +0.2  & 1.5  & 43 - 46 & 2.8 - 3.0 & 79  & 38\\
    GMF41.0-41.3  &  40.8 &  41.4 &  -0.5 &   +0.4  & 1.3  & 34 - 42 & 2.4 - 3.0 & 51 & -- \\
    GMF54.0-52.0  &  52.3 &  54.1 &  -0.3 &   +0.3  & 2.2  & 20 - 26 & 1.9 - 2.2 & 68 & 74 \\
    \hline
    Nessie$^2$ & 337.7 & 339.1 & -0.6 & -0.4 & 1.5 & 35 - 41  & (3.1)  & 81 & 74 \\
    G32.02+0.06$^3$ & 31.3 & 32.2 & -0.1 & 0.3 & 1.0 & 92 - 100 & 5.5 - 5.6 & 80 & 100 \\

\hline
  \end{tabular}\\
\end{center}
  \footnotesize{Column 1: GMF name; Column 2-5: upper and lower Galactic longitude and
  latitude; Column 6: Angular length of the filament from end to end in degrees;
  Column 7: Centroid velocity range spanned by filament in km s$^{-1}$;
  Column 8: Kinematic ``near'' distance computed using the \citet{Reid2009} model
  assuming standard Galactic parameters;
  Column 9: Projected length of filament;
  Column 10: Average velocity gradient along filament.  
  Comments: (1) This filament may extend to higher latitudes not probed by the current data.
  (2) From \citet{Jackson2010}. (3) From \citet{Battersby2012}. 
  }
\end{table*}

%% file: filaments_galactic_saga.tex
\begin{table*}[htbp]
  \caption{Physical properties of GMFs \label{tab:galactic}} 
\begin{center}
  \begin{tabular}{l c c c | c c c l}
    \hline
 Name & Cloud & Dense gas & DGMF  & $R_\mathrm{gal}$ & $\beta$ & $<z>$ & Assoc. \\
      &   mass & mass &  &   & & \\
      & [M$_{\odot}$] &  [M$_{\odot}$] & [\%] &  [kpc] & [degrees] & [pc] & \\
    \hline
    GMF18.0-16.8  &  1.5e+5 & 3.9e+3 &      2.7  &   6.3 &  4.6  & 55 & M16, W37 \\  
    GMF20.0-17.9  &  4.0e+5 & 4.8e+4 &     12.0  &  5.0 &  7.7  & (12) &  W39, SC-arm\\
    GMF26.7-25.4  &  2.0e+5 & 1.3e+4 &      6.5  &  5.7 &  9.4  & 68 & \\
    GMF38.1-32.4a &  7.0e+5 & 3.7e+4$^1$ &  5.3  &  5.9 & 12.8  & (24) & W44 \\
    GMF38.1-32.4b &  7.7e+4 & 5.0e+3$^1$ &  6.5  &   6.2 & 11.5  & (5) & \\
    GMF41.0-41.3  &  4.9e+4 & 7.7e+2 &      1.6  &   6.5 & 12.5  & 19 & \\
    GMF54.0-52.0  &  6.8e+4 & 2.4e+3 &      3.5  &  7.3 & 11.2  & 25 & W52\\
    \hline
    Nessie                       & --           & 3.9e+5 &   --    & 5.6 & -7.8 & -- & SC-arm \\
    G32.02+0.06 &  2.0e+5 & 3.0e+4 &     15.0  & 4.7 &  19.7 & 48 \\
\hline 
  \end{tabular}\\
\end{center}
  \footnotesize{Column 1: GMF name;
  Column 2: Total cloud derived from the integrated $^{13}$CO emission from the GRS {\bf down to 1\,K\,km\,s$^{-1}$}; 
  Column 3: Dense gas mass derived from dust continuum with ATLASGAL 870\micron data;
  Column 4: Dense gas mass fraction;
  Column 5: Galactocentric radius;
  Column 6: Angle measured from the Galactic centre-Sun line to the filament's position in the disk; 
  Column 7: Mean height above the Galactic midplane. Values in parentheses are for filaments which intersect with the projected plane.
  Column 8: Association with star formation region or spiral arm. \\
  Comments: (1) The dense gas mass from the two filaments
  within GMF\,38.1-32.4 partly overlap. While the envelope mass estimated
  from spectroscopic \dzco data accounts for the different velocity
  components, in the overlap regions, the ATLASGAL continuum data sums contributions from both components.  }
\end{table*}

%% file: filaments_images_saga.tex
\begin{figure*}[tbp]
\includegraphics[width=1.\textwidth]{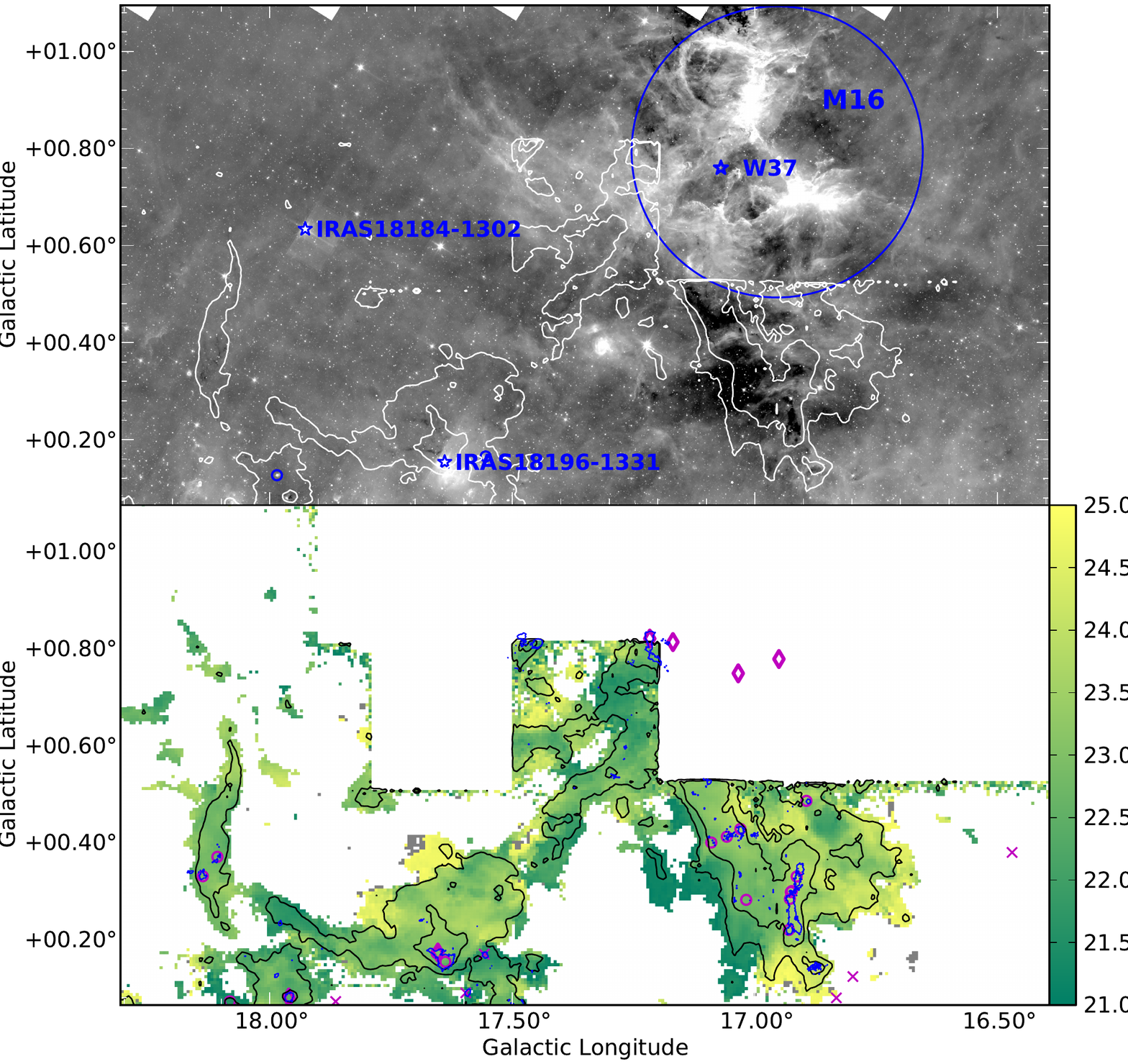}
  \caption{{\it Top:} Grayscale of the 8\,$\mu$m GLIMPSE image of GMF\,18.0-16.8. The white contour shows the $^{13}$CO integrated intensity of 2 and 7\,K\,km\,s$^{-1}$.  IRAS sources and W37 are marked with blue stars, and \hii regions found in the CORNISH survey are shown in blue circles.  M16 is a prominent emission source in this region, north of the filament but at a consistent characteristic velocity.  
{\it Bottom:} Colourscale of the centroid velocity field from the GRS $^{13}$CO data where the integrated intensity is above 1\,K\,km\,s$^{-1}$ in the indicated velocity range, shown in the colour bar in km s$^{-1}$. The blue contour is ATLASGAL 870$\mu$m emission of 250\,mJy beam$^{-1}$.  In magenta circles, we show all $v_\mathrm{lsr}$ measurements of BGPS clumps that have velocities within the $^{13}$CO velocity range and with $\times$ those $v_\mathrm{lsr}$ values outside of the $^{13}$CO velocity range. The magenta diamonds are positions in the \cite{Wienen2012} catalog with a matching velocity. The black contours are the $^{13}$CO integrated intensity at 2 and 7\,K\,km\,s$^{-1}$.}
  \label{fig:filament_16}
\end{figure*}

\begin{figure*}[tbp]
  \includegraphics[width=1.\textwidth]{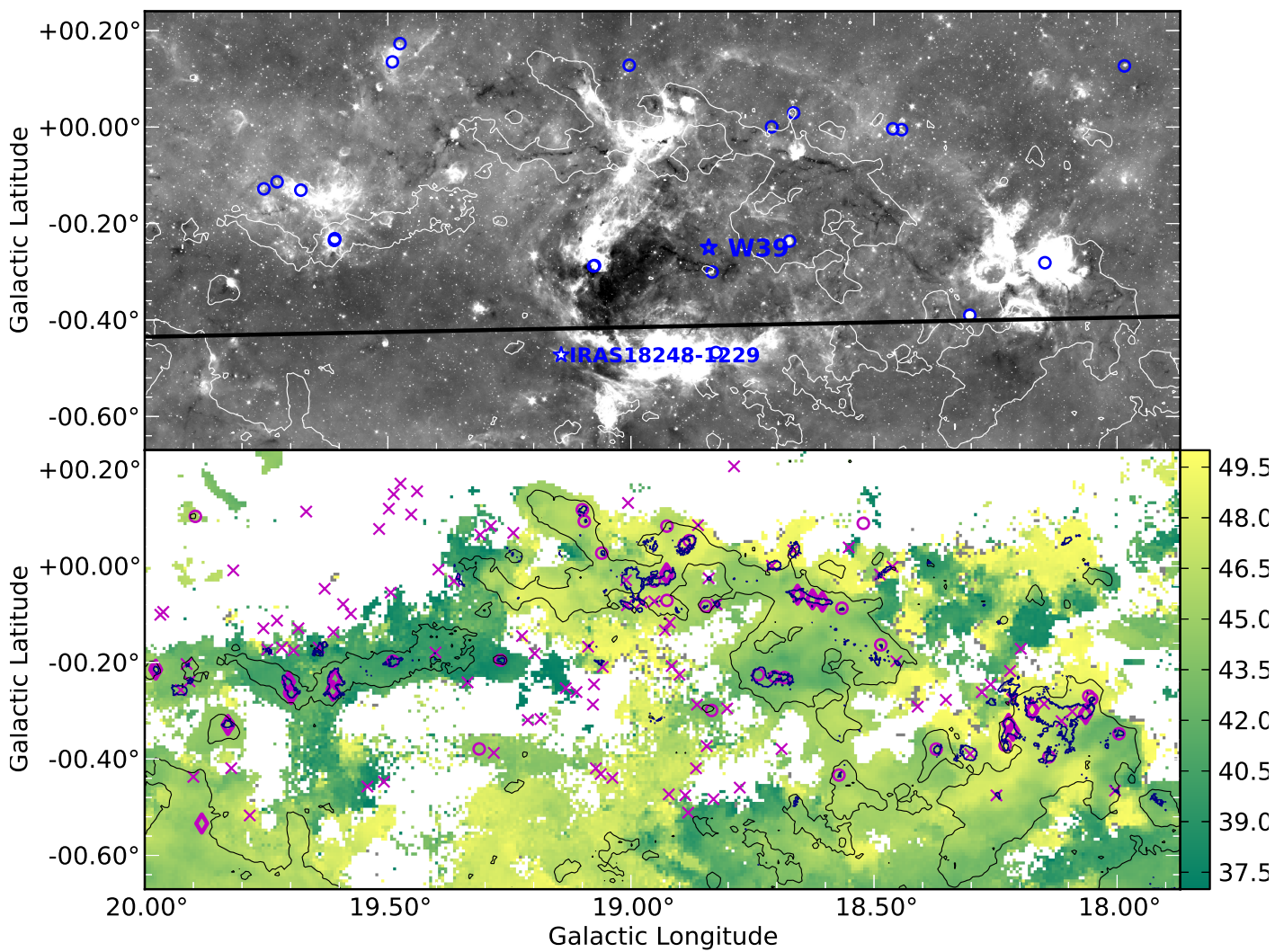}
  \caption{{\it Top:} Grayscale of the 8\,$\mu$m GLIMPSE image of GMF\,20.0-17.9. The white contour shows the $^{13}$CO integrated intensity of 3.5\,K\,km\,s$^{-1}$.  IRAS\,18248-1229 and W39 are both labeled with blue stars, and \hii regions from the CORNISH survey are marked with blue circles.  The black line is the location of the Galactic plane at the distance of this filament. 
{\it Bottom:} Colourscale of the centroid velocity field from the GRS $^{13}$CO data where the integrated intensity is above 1\,K\,km\,s$^{-1}$ in the indicated velocity range, shown in the colour bar in km s$^{-1}$. The blue contour is ATLASGAL 870$\mu$m emission of 250\,mJy beam$^{-1}$..  In magenta circles, we show all $v_\mathrm{lsr}$ measurements of BGPS clumps that have velocities within the $^{13}$CO velocity range and with $\times$ those $v_\mathrm{lsr}$ values outside of the $^{13}$CO velocity range. The diamonds are positions in the \cite{Wienen2012} catalog with a matching velocity.  The black contour is the $^{13}$CO integrated intensity at 3.5 K\,km\,s$^{-1}$. }
  \label{fig:filament_17}
\end{figure*}

\begin{figure*}[tbp]
  \includegraphics[width=1.\textwidth]{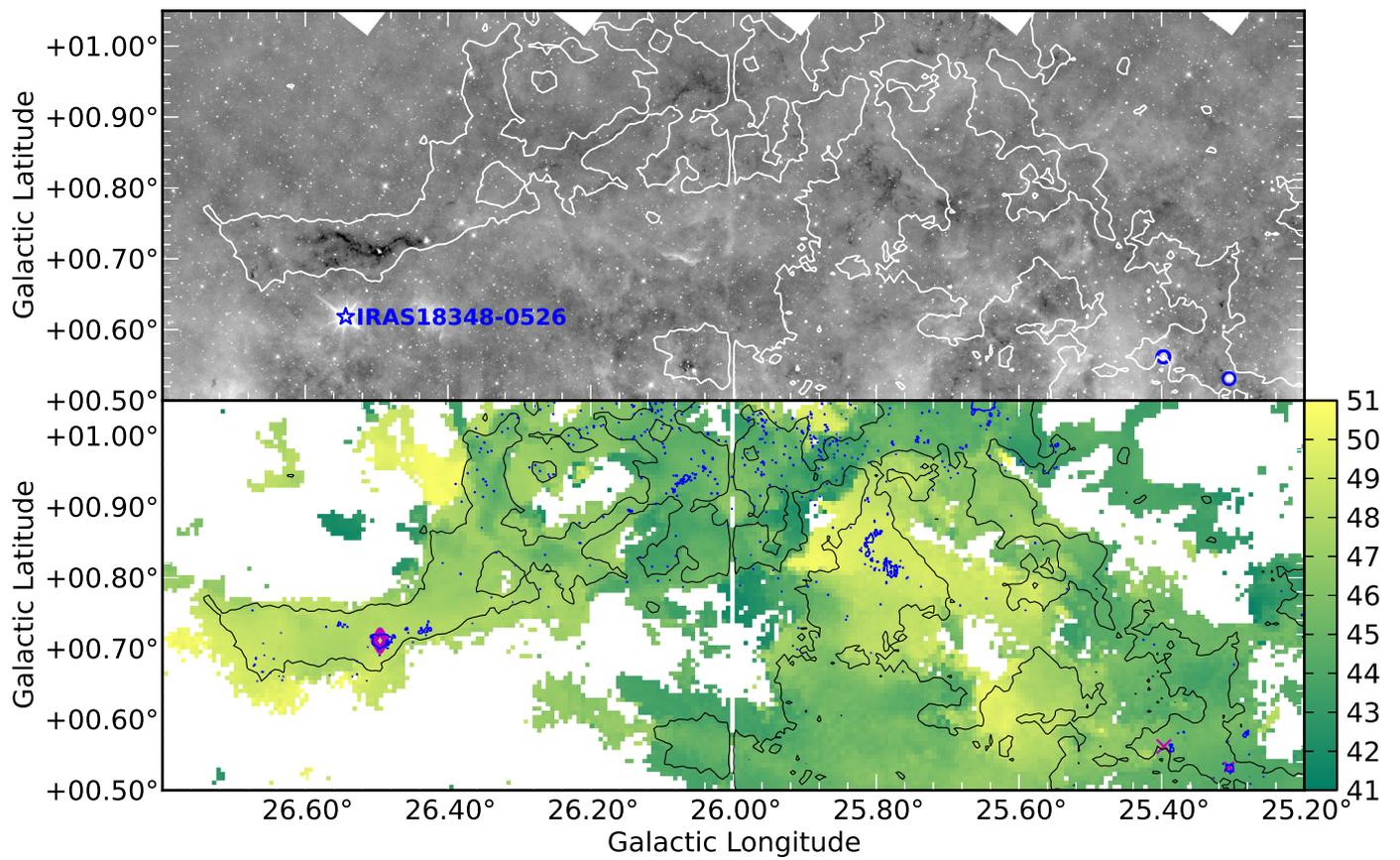}
  \caption{{\it Top:} Grayscale of the 8\,$\mu$m GLIMPSE image of GMF\,26.7-25.4. The white contour shows the $^{13}$CO integrated intensity of 2\,K\,km\,s$^{-1}$.  The blue star is the position of IRAS\,18348-0526, and the blue circles are CORNISH detected \hii regions. 
{\it Bottom:} Colourscale of the centroid velocity field from the GRS $^{13}$CO data where the integrated intensity is above 1\,K\,km\,s$^{-1}$ in the indicated velocity range, shown in the colour bar in km s$^{-1}$. The blue contour is ATLASGAL 870$\mu$m emission of 250\,mJy beam$^{-1}$.  The magenta diamond indicates a pointing from the \citet{Wienen2012} NH$_3$ catalog that has a consistent velocity, though the other two pointings ($\times$, in the bottom right corner) have velocities outside of the $^{13}$CO range. The black contour is the $^{13}$CO integrated intensity at 2\,K\,km\,s$^{-1}$. 
}
  \label{fig:filament_26}
\end{figure*}

\begin{figure*}[tbp]
 \includegraphics[angle=-90, width=1.\textwidth]{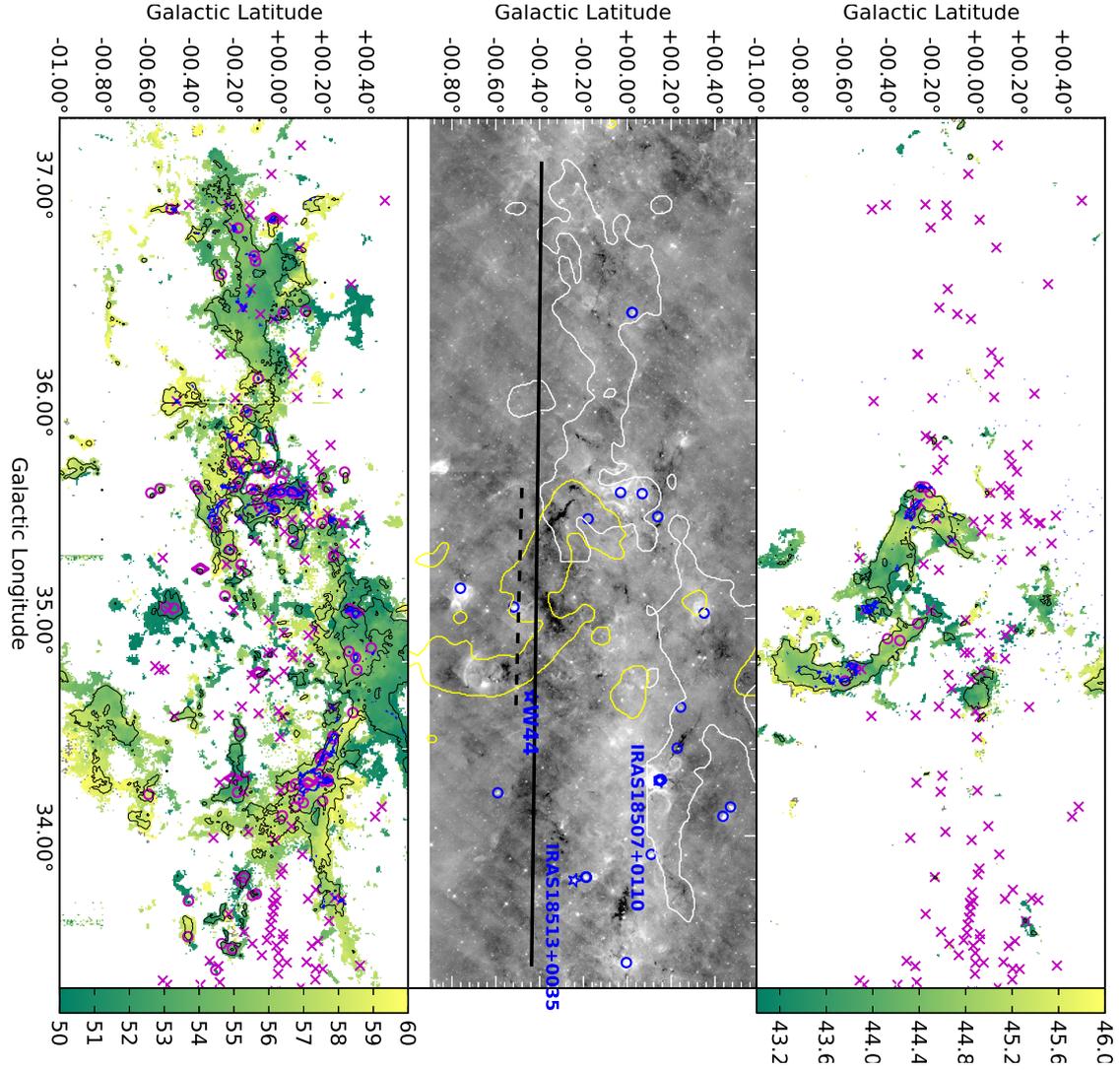}
  \caption{{\it Centre:} Grayscale of the 8\,$\mu$m GLIMPSE image of GMF\,38.1-32.4. The white contour shows the $^{13}$CO  integrated intensity of 0.5\,K\,km\,s$^{-1}$ between 50 and 60\,km s$^{-1}$ (bottom), and the yellow contour shows the $^{13}$CO integrated intensity of 0.5\,K\,km\,s$^{-1}$ between 43 and 46\,km s$^{-1}$ (top).  Blue stars indicate the positions of IRAS sources and W44, and blue circles are the positions of \hii regions from the CORNISH survey.  The solid black line is the location of the Galactic plane at the distance of the 50-60\,km s$^{-1}$ filament, and the dashed black line corresponds to the plane at the distance of the 43-46\,km s$^{-1}$. 
{\it Top and bottom:} Colourscale of the centroid velocity field from the GRS $^{13}$CO data for the 43 to 46 km s$^{-1}$ (top) and 50 to 60 km s$^{-1}$ (bottom) filaments where the integrated intensity is above 1\,K\,km\,s$^{-1}$ in the indicated velocity range, shown in the colour bar in km s$^{-1}$.  The blue contour is ATLASGAL 870$\mu$m emission of 250\,mJy beam$^{-1}$ within each masked region.  In magenta circles, we show all $v_\mathrm{lsr}$ measurements of BGPS clumps that have velocities within the respective $^{13}$CO velocity ranges and with $\times$ those $v_\mathrm{lsr}$ values outside of the $^{13}$CO velocity ranges. The diamonds are positions in the \cite{Wienen2012} catalog with a matching velocity. The black contours are the $^{13}$CO integrated intensity of 2 and 7\,K\,km\,s$^{-1}$.  }
  \label{fig:filament_38}
\end{figure*}

\begin{figure*}[tbp]
  \includegraphics[width=1.\textwidth]{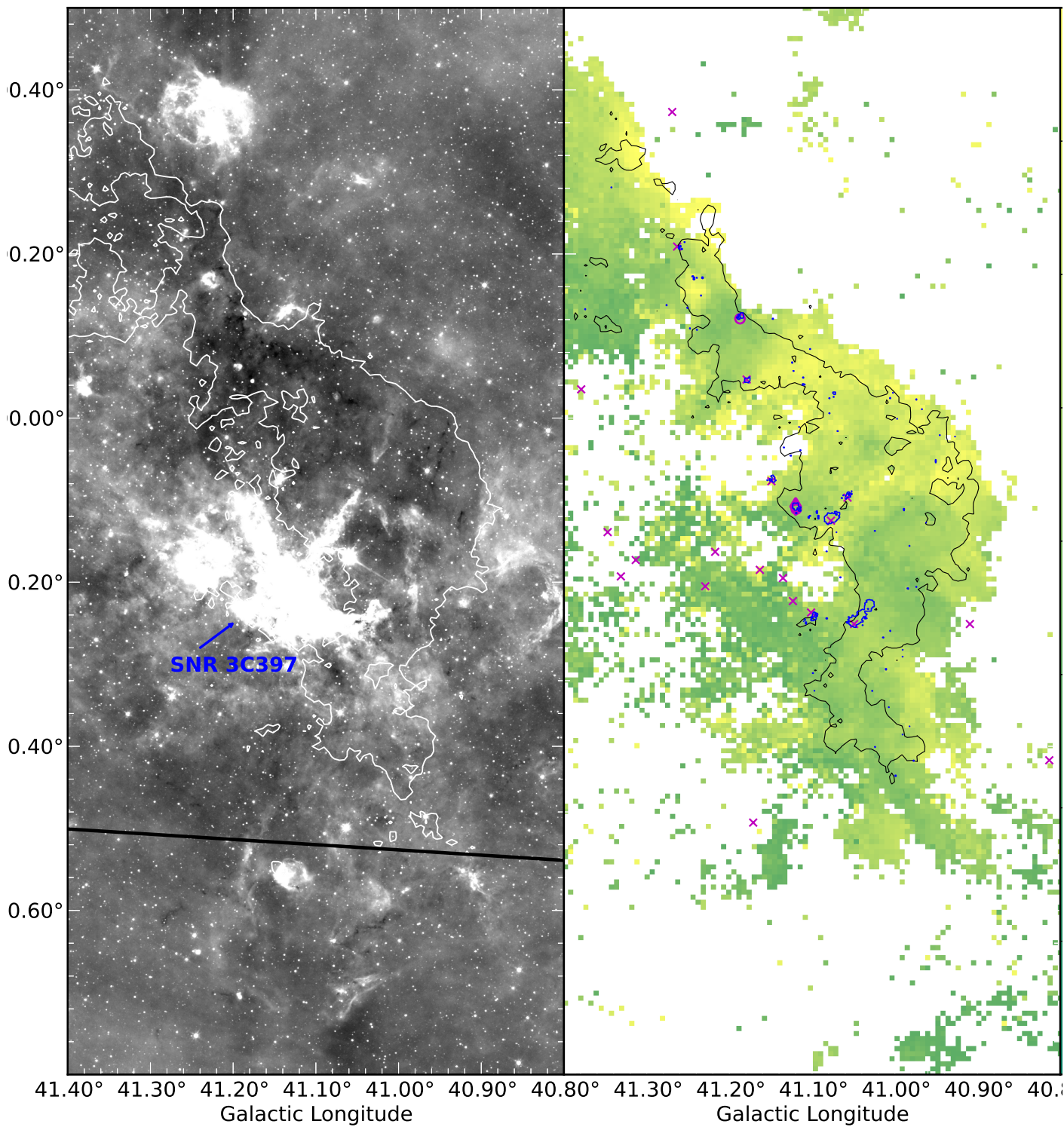}
  \caption{{\it Left:} Grayscale of the 8\,$\mu$m GLIMPSE image of GMF\,41.0-41.3. The white contour shows the ${13}$CO integrated intensity of 1\,K\,km\,s$^{-1}$.  The solid black line shows the location of the Galactic midplane at the distance of this filament. The solid black line is the location of the Galactic plane at the distance of this filament.  The supernova remnant 3C397 is the (unrelated) bright IR emission region.  {\it Right:} Colourscale of the centroid velocity field from the GRS $^{13}$CO data where the integrated intensity is above 1\,K\,km\,s$^{-1}$ in the indicated velocity range, shown in the colour bar in km s$^{-1}$. The blue contour is ATLASGAL 870$\mu$m emission of 250\,mJy beam$^{-1}$.  In magenta circles, we show all $v_\mathrm{lsr}$ measurements of BGPS clumps that have velocities within the $^{13}$CO velocity range and with $\times$ those $v_\mathrm{lsr}$ values outside of the $^{13}$CO velocity range. The diamonds are positions in the \cite{Wienen2012} catalog with a matching 
velocity. The black contours are the $^{13}$CO integrated intensity of 2 and 7\,K\,km\,s$^{-1}$.  }
  \label{fig:filament_41}
\end{figure*}

\begin{figure*}[tbp]
\includegraphics[width=1.\textwidth]{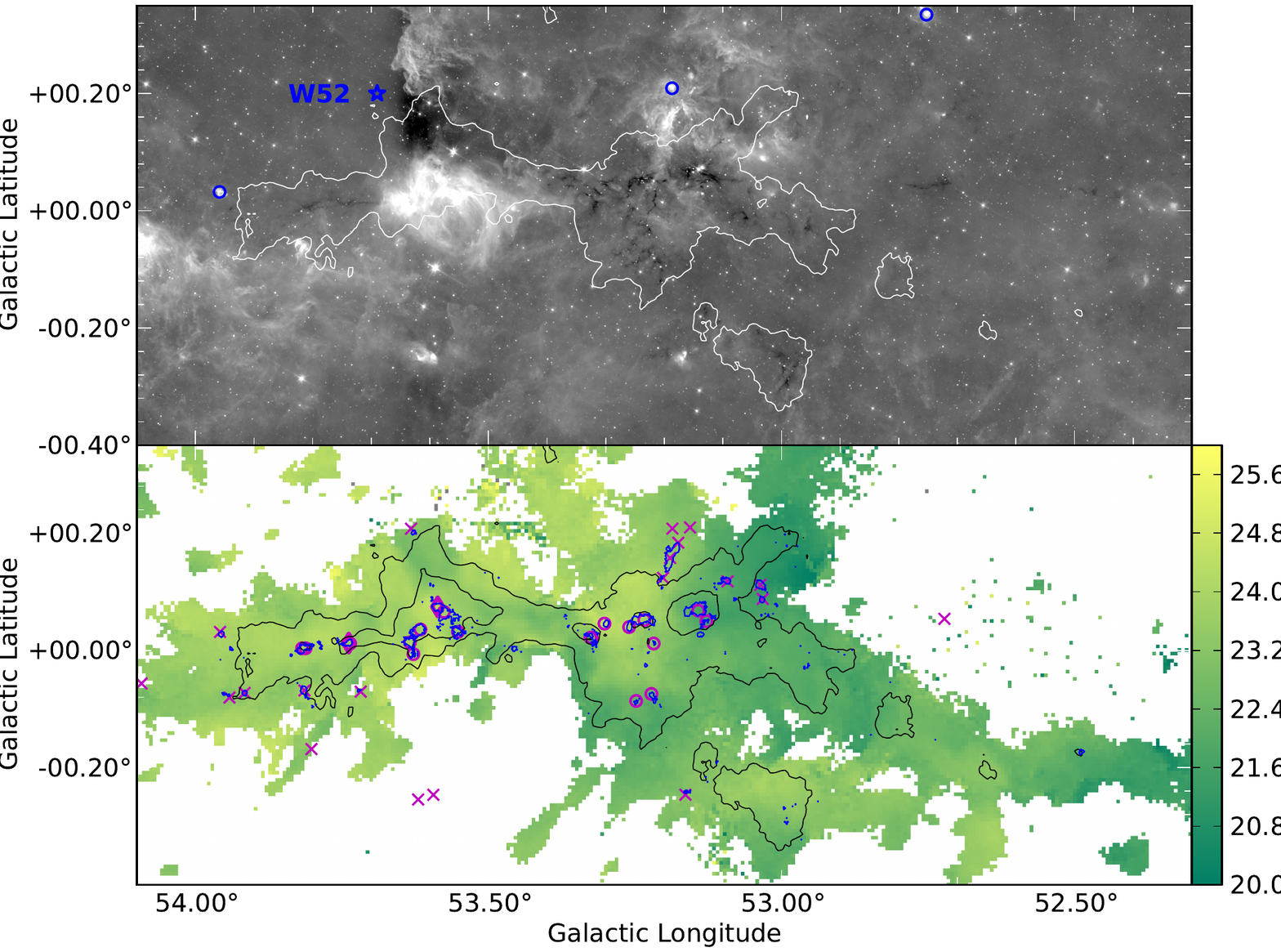}
  \caption{{\it Top:} Grayscale of the 8\,$\mu$m GLIMPSE image of GMF\,54.0-52.0. The white contour shows the ${13}$CO integrated intensity of 2\,K\,km\,s$^{-1}$. The blue star shows the location of W52, and the blue circles show CORNISH detections of \hii regions. 
{\it Bottom:} Colourscale of the centroid velocity field from the GRS $^{13}$CO data where the integrated intensity is above 1\,K\,km\,s$^{-1}$ in the indicated velocity range, shown in the colour bar in km s$^{-1}$. The blue contour is ATLASGAL 870$\mu$m emission of 250\,mJy beam$^{-1}$.  In symbols, we show all $v_\mathrm{lsr}$ measurements of BGPS clumps, indicating in circles those clumps that have velocities within the $^{13}$CO velocity range and with $\times$ those $v_\mathrm{lsr}$ values outside of the $^{13}$CO velocity range. The diamonds are positions in the \cite{Wienen2012} catalog with a matching velocity.  The black contours are the $^{13}$CO integrated intensity at 2 and 7 K\,km\,$^{-1}$.}
  \label{fig:filament_54}
\end{figure*}

%% file: filaments_pv_saga.tex
\clearpage
\begin{figure}
\includegraphics[width=1.0\textwidth]{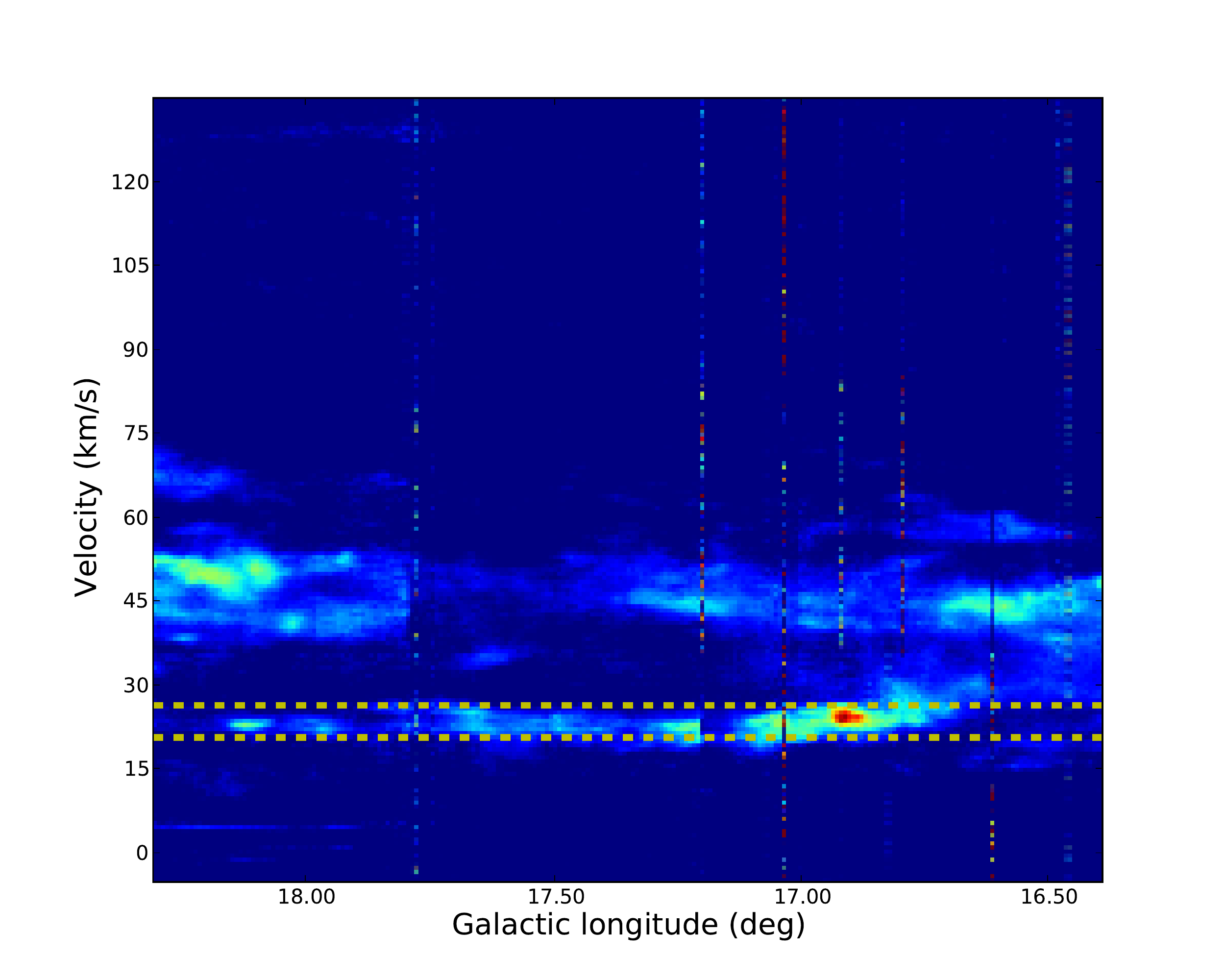}
\caption{Position velocity diagram for the longitude range of GMF\,18.0-16.8. The selected velocity range (21 - 25\,km\,s$^{-1}$) is outlined in the yellow dashed lines.}
\label{fig:fil16_pv}
\end{figure}

\clearpage
\begin{figure}
\includegraphics[width=1.0\textwidth]{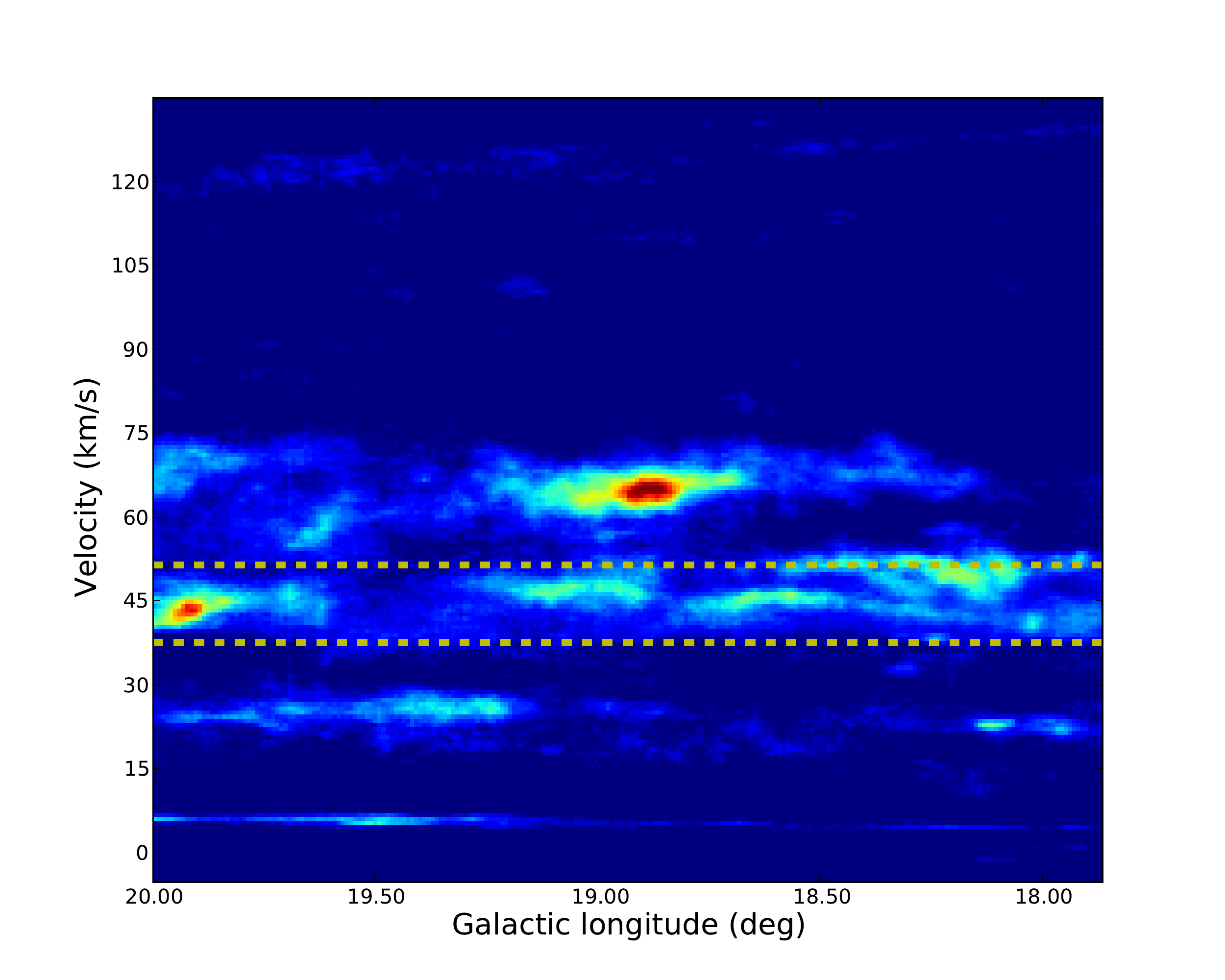}
\caption{Position velocity diagram for the longitude range of GMF\,20.0-17.9. The selected velocity range (37 - 50\,km\,s$^{-1}$) is outlined in the yellow dashed lines.}
\label{fig:fil17_pv}
\end{figure}

\clearpage
\begin{figure}
\includegraphics[width=1.0\textwidth]{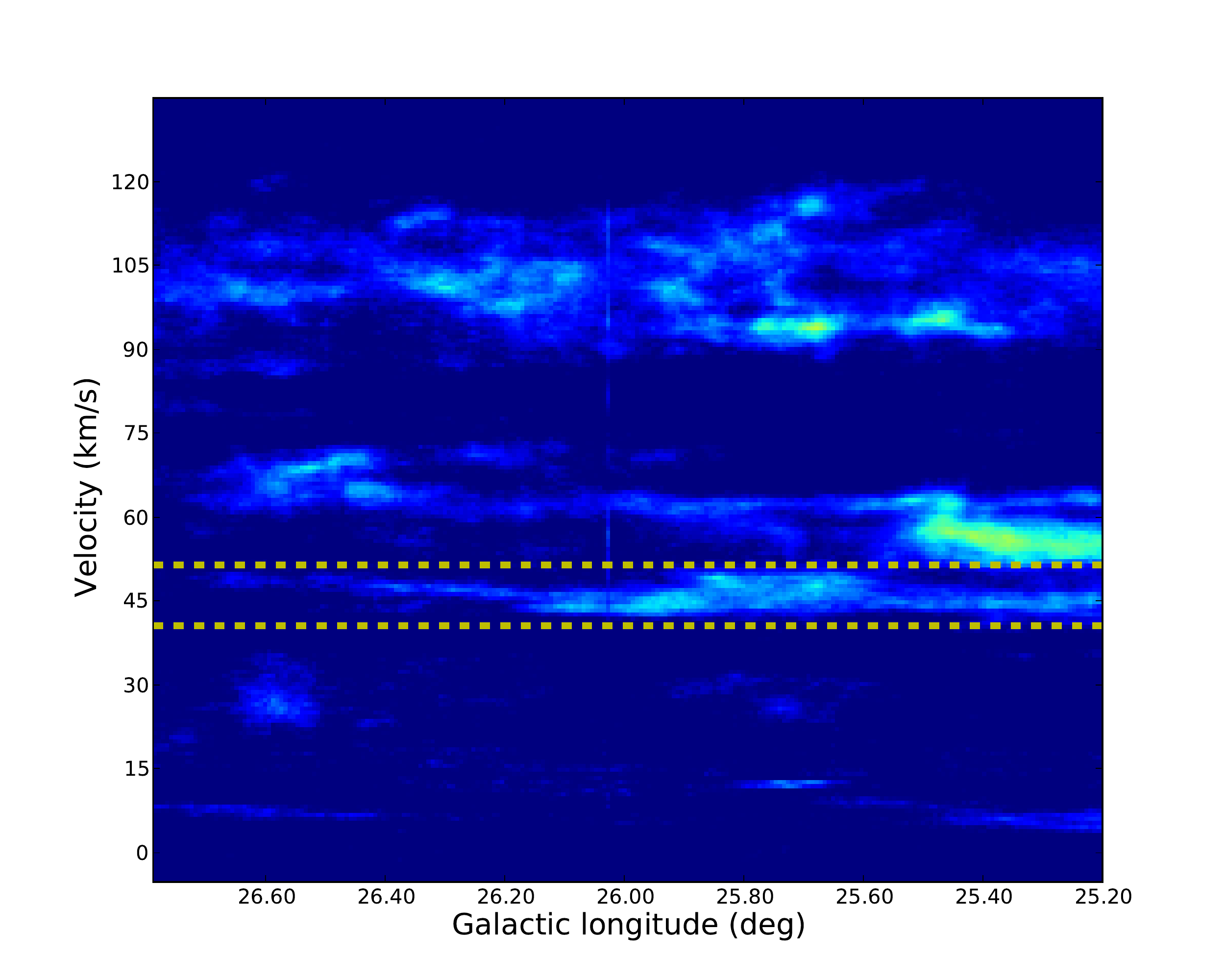}
\caption{Position velocity diagram for the longitude range of GMF\,26.7-25.4. The selected velocity range (41 - 51\,km\,s$^{-1}$) is outlined in the yellow dashed lines.}
\label{fig:fil26_pv}
\end{figure}

\clearpage
\begin{figure}
\includegraphics[width=1.0\textwidth]{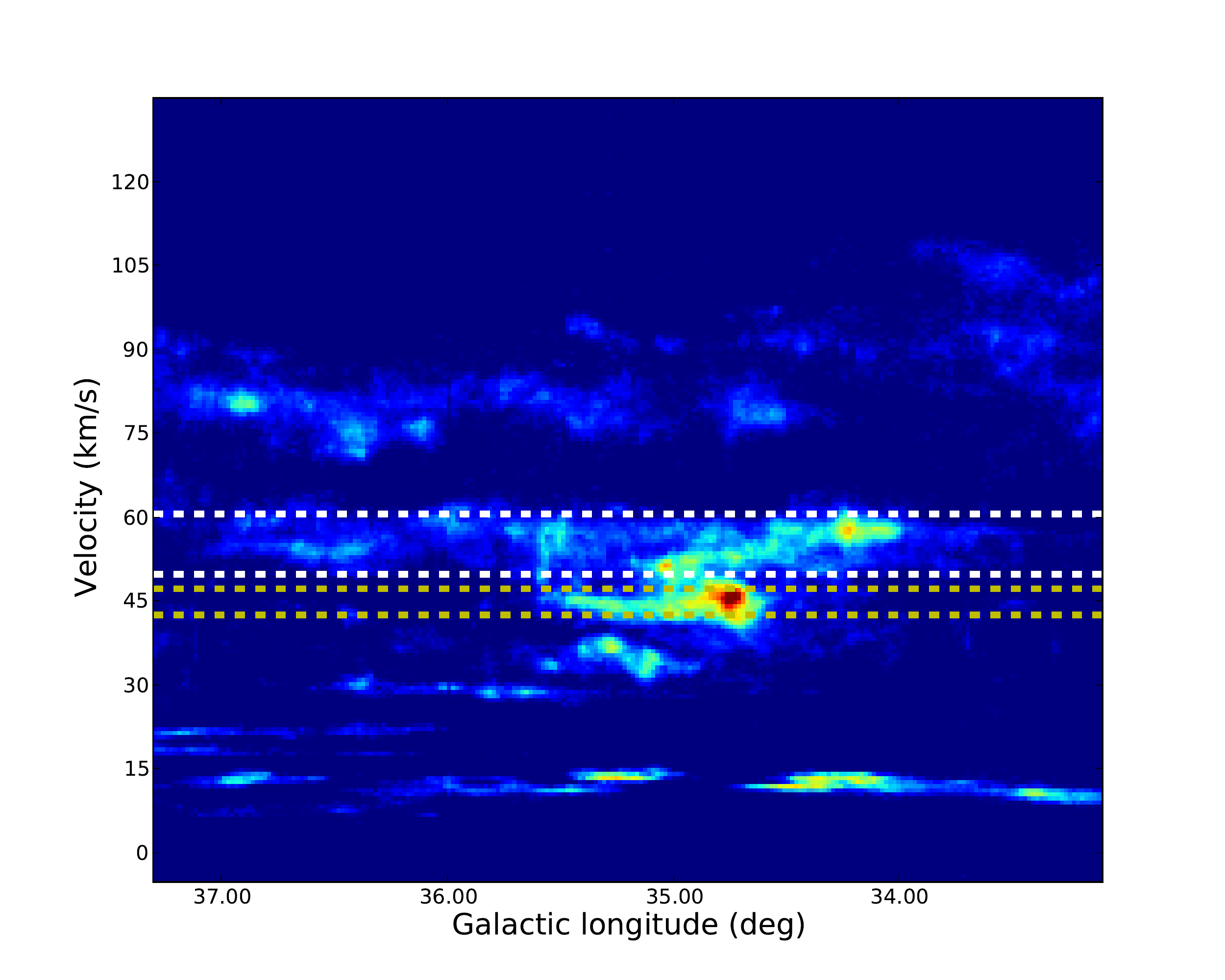}
\caption{Position velocity diagram for the longitude range of GMF\,38.1-32.4. The selected velocity range is outlined in the white (50 - 60\,km\,s$^{-1}$) and yellow (43 - 46\,km\,s$^{-1}$) dashed lines.}
\label{fig:fil38_pv}
\end{figure}

\clearpage
\begin{figure}
\includegraphics[width=1.0\textwidth]{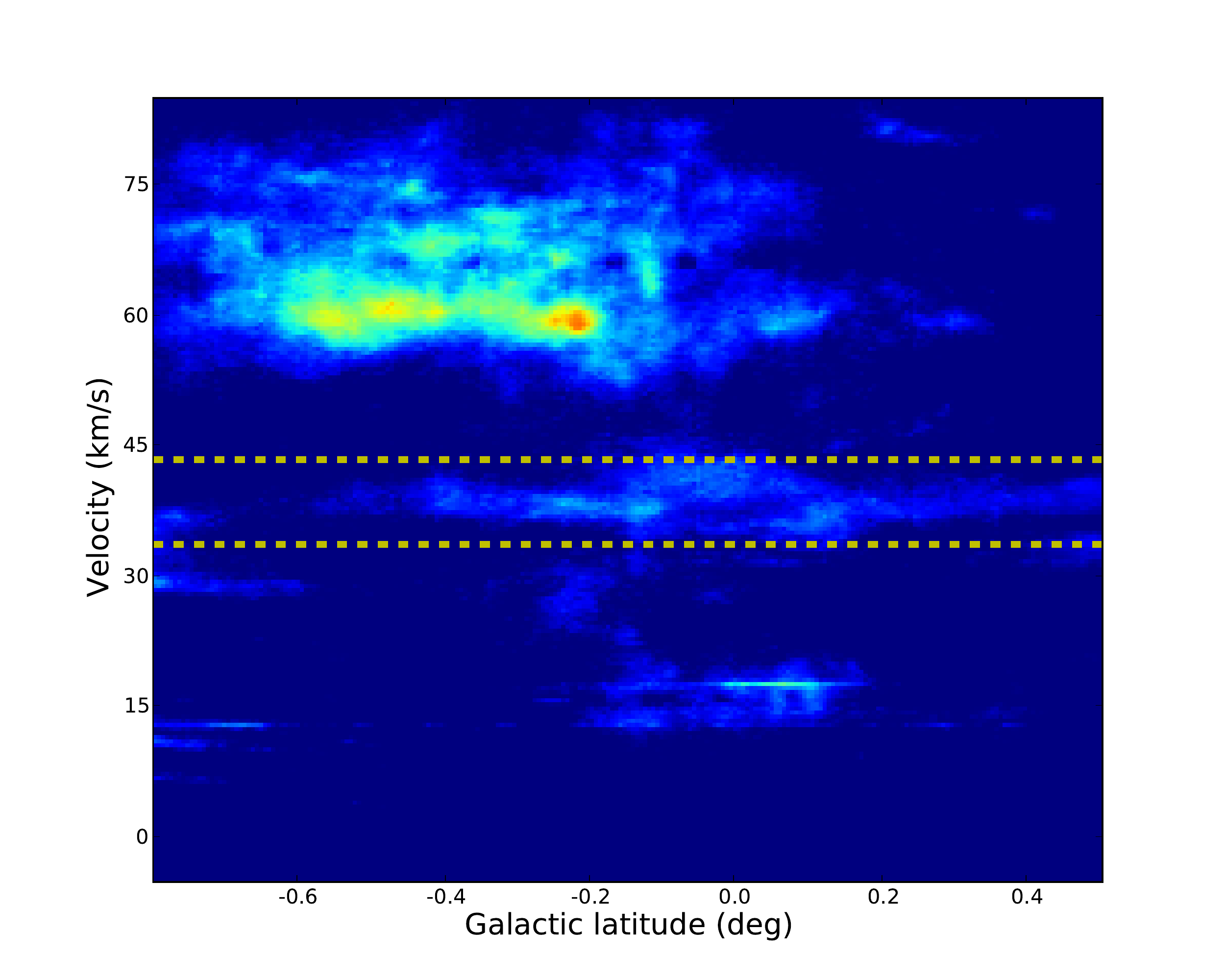}
\caption{Position velocity diagram for the latitude range of vertical GMF\,41.0-41.3. The selected velocity range (34 - 42\,km\,s$^{-1}$) is outlined in the yellow dashed lines.}
\label{fig:fil41_pv}
\end{figure}

\clearpage
\begin{figure}
\includegraphics[width=1.0\textwidth]{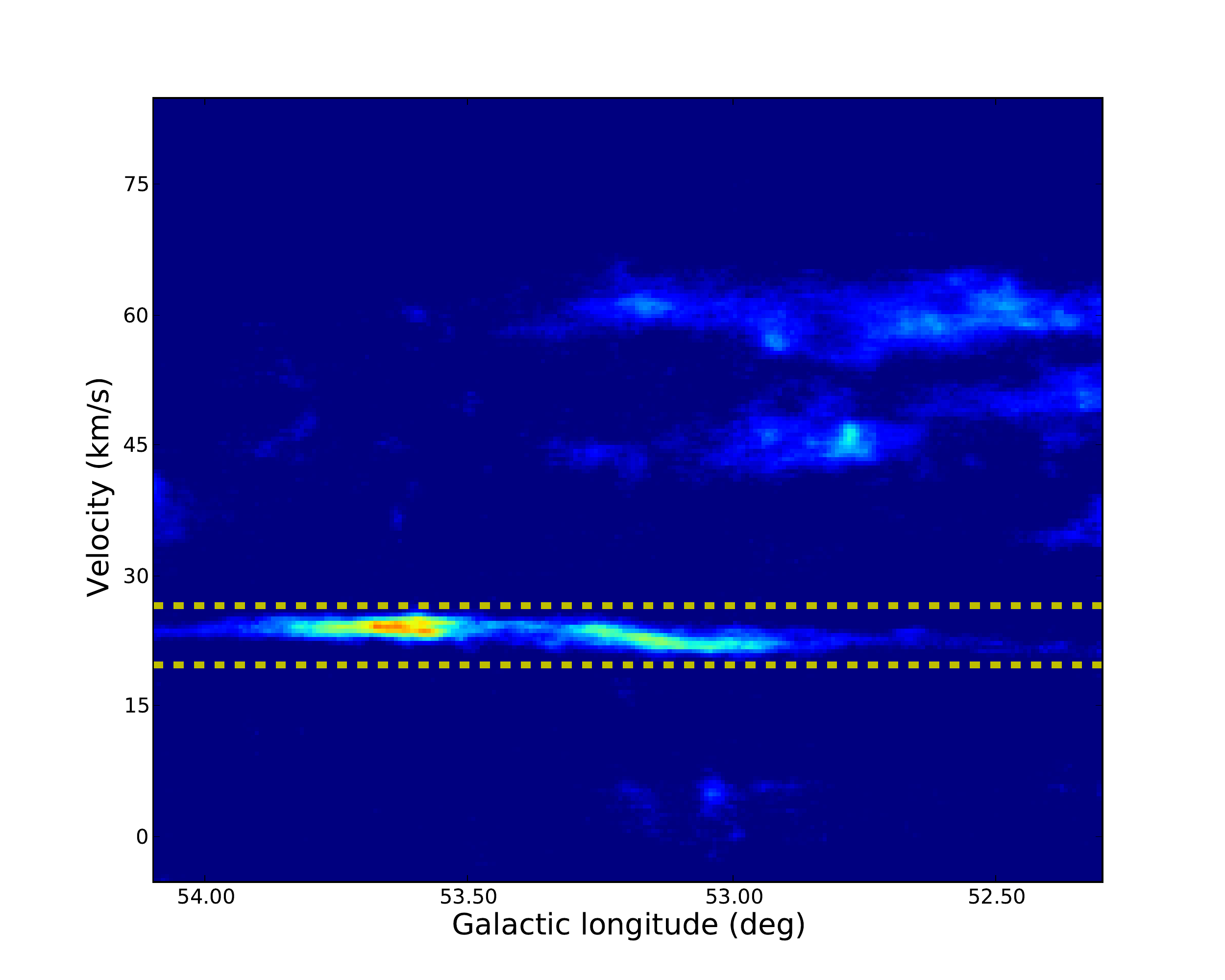}
\caption{Position velocity diagram for the longitude range of GMF\,54.0-52.0. The selected velocity range (20 - 26\,km\,s$^{-1}$) is outlined in the yellow dashed lines.}
\label{fig:fil54_pv}
\end{figure}